%% file: main.tex
\documentclass[sigconf]{acmart}

\AtBeginDocument{%
  }

\usepackage{enumitem}
\usepackage{amsmath}
\usepackage{amsthm}          
\usepackage{subcaption}
\usepackage{xspace}
\usepackage{multirow}
\usepackage{makecell}
\usepackage[htt]{hyphenat}    
\usepackage{algorithm}
\usepackage{algpseudocode}

\newcommand{\sysname}{HKV\xspace}

\setcopyright{acmlicensed}
\copyrightyear{2027}
\acmYear{2027}
\acmDOI{XXXXXXX.XXXXXXX}

\acmConference[SIGMOD '27]{Proceedings of the 2027 International Conference
  on Management of Data}{June 22--27, 2027}{Berlin, Germany}
\acmBooktitle{Proceedings of the 2027 International Conference on Management
  of Data (SIGMOD '27), June 22--27, 2027, Berlin, Germany}
\acmISBN{978-1-4503-XXXX-X/2027/06}

\begin{document}

\title{HierarchicalKV: A GPU Hash Table with Cache Semantics for Continuous Online Embedding Storage}

\author{Haidong Rong}
\authornote{Corresponding author. Equal contribution.}
\orcid{0009-0003-2103-0241}
\affiliation{%
  \institution{NVIDIA}
  \country{USA}
}
\email{hrong@nvidia.com}

\author{Jiashu Yao}
\authornote{Equal contribution.}
\orcid{0009-0008-7129-1294}
\affiliation{%
  \institution{NVIDIA}
  \country{China}
}
\email{jiashuy@nvidia.com}

\author{Matthias Langer}
\orcid{0000-0003-1776-8000}
\affiliation{%
  \institution{NVIDIA}
  \country{Australia}
}
\email{mlanger@nvidia.com}

\author{Shijie Liu}
\orcid{0009-0007-5431-5787}
\affiliation{%
  \institution{NVIDIA}
  \country{China}
}
\email{aleliu@nvidia.com}

\author{Li Fan}
\orcid{0009-0004-4306-5872}
\affiliation{%
  \institution{Tencent}
  \country{China}
}
\email{oppenheimli@tencent.com}

\author{Dongxin Wang}
\orcid{0009-0001-2403-2977}
\affiliation{%
  \institution{Vipshop}
  \country{China}
}
\email{dxwang.bill@gmail.com}

\author{Jia He}
\orcid{0000-0003-2513-1595}
\affiliation{%
  \institution{BOSS Zhipin}
  \country{China}
}
\email{hejia01@kanzhun.com}

\author{Jinglin Chen}
\orcid{0009-0000-6912-1837}
\affiliation{%
  \institution{BOSS Zhipin}
  \country{China}
}
\email{chenjinglin@kanzhun.com}

\author{Jiaheng Rang}
\orcid{0009-0003-9677-4886}
\affiliation{%
  \institution{ByteDance}
  \country{China}
}
\email{rangjiaheng@gmail.com}

\author{Julian Qian}
\affiliation{%
  \institution{Snap}
  \country{USA}
}
\email{julian.qian@gmail.com}

\author{Mengyao Xu}
\orcid{0009-0000-5772-1739}
\affiliation{%
  \institution{NVIDIA}
  \country{USA}
}
\email{mengyaox@nvidia.com}

\author{Fan Yu}
\affiliation{%
  \institution{NVIDIA}
  \country{China}
}
\email{fayu@nvidia.com}

\author{Minseok Lee}
\orcid{0000-0002-8367-1939}
\affiliation{%
  \institution{NVIDIA}
  \country{USA}
}
\email{minseokl@nvidia.com}

\author{Zehuan Wang}
\orcid{0000-0002-1072-2651}
\affiliation{%
  \institution{NVIDIA}
  \country{China}
}
\email{zehuanw@nvidia.com}

\author{Even Oldridge}
\affiliation{%
  \institution{NVIDIA}
  \country{Canada}
}
\email{eoldridge@nvidia.com}

\renewcommand{\shortauthors}{Rong and Yao, et al.}

\input{sections/abstract}

\ccsdesc[500]{Information systems~Data structures}
\ccsdesc[300]{Computer systems organization~Parallel architectures}
\ccsdesc[500]{Information systems~Hash table design}

\keywords{GPU, hash table, cache semantics, eviction, embedding storage, recommendation systems, key-value store}

\maketitle

\input{sections/s1-introduction}
\input{sections/s2-background}
\input{sections/s3-design}
\input{sections/s4-implementation}
\input{sections/s5-evaluation}
\input{sections/s6-related}
\input{sections/s7-conclusion}

\begin{acks}
We thank Yafei Zhang (Tencent) for his valuable contributions to the early design, implementation, and code review of HKV during the formative stages of this project.
\end{acks}

\bibliographystyle{ACM-Reference-Format}
\bibliography{references}

\end{document}

%% file: sections/abstract.tex
\begin{abstract}
Traditional GPU hash tables preserve every inserted key---a dictionary assumption that wastes scarce High Bandwidth Memory (HBM) when embedding tables routinely exceed single-GPU capacity.  We challenge this assumption with \emph{cache semantics}, where policy-driven eviction is a first-class operation.
We introduce HierarchicalKV (\sysname), the first general-purpose GPU hash table library whose normal full-capacity operating contract is cache-semantic: each full-bucket upsert (update-or-insert) is resolved in place by eviction or admission rejection rather than by rehashing or capacity-induced failure.  \sysname co-designs four core mechanisms---cache-line-aligned buckets, in-line score-driven upsert, score-based dynamic dual-bucket selection, and triple-group concurrency---and uses tiered key-value separation as a scaling enabler beyond HBM.
On an NVIDIA H100 NVL GPU, \sysname achieves up to
3.9 billion key--value pairs per second (B-KV/s) \texttt{find} throughput, stable across load factors
0.50--1.00 ({<}5\% variation), and delivers
1.4$\times$ higher \texttt{find} throughput than WarpCore (the strongest dictionary-semantic GPU baseline at $\lambda{=}0.50$) and up to 2.6--9.4$\times$ over indirection-based GPU baselines.
Since its open-source release in October~2022, \sysname has been integrated into multiple open-source recommendation frameworks.
\end{abstract}

%% file: sections/s1-introduction.tex
\section{Introduction}
\label{sec:intro}

\looseness=-1
Modern recommendation, search,
and advertising systems~\cite{naumov2019dlrm} rely on deep learning
models whose \emph{embedding tables}---mapping
sparse features to dense vectors---are the dominant memory consumer,
routinely exceeding single-GPU HBM capacity (e.g., a billion
64-dim \texttt{float32} keys occupy ${\sim}$256\,GB vs.\ 94\,GB on
an NVIDIA~H100 NVL).  Continuous online training further intensifies
the pressure, demanding sustained key ingestion under a hard memory
budget.

\looseness=-1
Every widely-used GPU hash table---cuCollections~\cite{nvidia2020cucollections},
WarpCore~\cite{junger2020warpcore},
BGHT~\cite{awad2022better}, WarpSpeed~\cite{mccoy2025warpspeed},
and Hive~\cite{polak2025hive}---adopts \emph{dictionary semantics}:
every inserted key must be preserved.  As the load factor approaches~1.0,
probe chains lengthen, throughput degrades, and the system must rehash
or fail.  These limitations are intrinsic to dictionary semantics:
as long as every key must be preserved, high-load-factor degradation
and external maintenance remain unavoidable.
Our measurements show that find throughput drops by
31--100\% as $\lambda$ approaches~1.0
(\S\ref{ssec:exp-lf}), rendering dictionary-semantic hash tables
unsuitable for continuously filled embedding caches.

\looseness=-1
We observe that embedding storage is better modeled as a
\emph{cache} than a dictionary: power-law access patterns mean
retaining high-value entries and evicting low-value ones preserves
model quality, while the hard memory budget makes eviction an
inevitable, recurring event.
This shift from dictionary to cache semantics fundamentally changes the design constraints: a full table is no longer an error state but the \emph{normal operating condition}.
Just as the distinction between B-tree indexes (dictionary semantics) and buffer pools (cache semantics) drives fundamentally different concurrency and replacement designs in database systems, the same semantic divide drives fundamentally different GPU hash table designs.
Embracing \emph{cache semantics} opens a new design space but
introduces four concrete challenges (\S\ref{ssec:challenges}):
bounding per-key lookup work to a fixed number of GPU memory
transactions~(C1), resolving full-bucket upserts in place without
rehashing~(C2), preserving high-value entries under bounded
associativity~(C3), and separating structural from
non-structural writes under mixed workloads~(C4).

\looseness=-1
To address these challenges, we introduce HierarchicalKV (hereafter \sysname), the first general-purpose GPU hash table \emph{library} whose normal full-capacity operating contract is cache-semantic: each full-bucket upsert (update-or-insert) is resolved in place by eviction or admission rejection, without rehashing or external maintenance.  \sysname synthesizes four co-designed mechanisms into a cohesive architecture:
\begin{enumerate}[nosep,leftmargin=*]
  \item \textbf{Single-bucket-confined cache-line-aligned bucket} (\S\ref{ssec:bucket}):
    single-bucket confinement collapses each key's entire candidate
    space into 128~one-byte digests occupying one GPU L1 cache line,
    enabling \emph{definitive per-bucket miss in a single memory
    transaction} (one transaction in single-bucket mode; two in
    dual-bucket mode).
  \item \textbf{In-line score-driven upsert semantics} (\S\ref{ssec:eviction}):
    bucket-local score comparison, admission control, and Compare-And-Swap (CAS) commit are
    fused directly into the insert path, so a full-bucket upsert is
    resolved inside the table with no separate eviction workflow.
  \item \textbf{Score-based dynamic dual-bucket selection}
    (\S\ref{ssec:dualbucket}): extends the power-of-two-choices (P2C) paradigm~\cite{azar1994balanced,mitzenmacher2001power} from load balancing to eviction-quality optimization in GPU hash tables, recovering the
    ${\sim}34\%$ HBM left unused when birthday-paradox collisions
    trigger premature eviction at $\lambda{\approx}0.66$ in
    single-bucket mode, and achieving 99.4\% top-$N$ score retention at
    $\lambda{=}1.0$.
  \item \textbf{Triple-group concurrency} (\S\ref{ssec:triplegroup}):
    reader/updater/inserter role separation with a CPU--GPU
    dual-layer lock, confining all structural modifications to
    the inserter kernel so that reader and updater kernels require
    no CAS or eviction logic---substantially reducing per-kernel
    complexity and enabling independent optimization of each
    access pattern.
\end{enumerate}
As a scaling enabler,
\emph{tiered key-value separation} (\S\ref{ssec:kvsep}) extends capacity
beyond HBM: keys, digests, and scores reside in HBM while values overflow
to pinned host memory (HMEM) via position-based addressing.  Together the
four core mechanisms are co-designed so that their interactions deliver
three properties absent from all prior GPU hash tables: fixed per-bucket
miss work at every load factor, full-capacity in-place upsert without
rehashing, and concurrent read/update/insert execution with role-isolated
kernels; tiered KV separation extends this design
to hybrid HBM+HMEM deployment while keeping key-side processing on GPU.

\looseness=-1
Our evaluation on an NVIDIA~H100 NVL GPU shows that \sysname
achieves up to 3.9\,B-KV/s \texttt{find} throughput, stable
across load factors 0.50--1.00 ({<}5\% variation), and delivers
1.4$\times$ the find throughput of WarpCore and up to 2.6--9.4$\times$ over indirection-based designs (i.e., those using (key,\,index) pairs with separate value gather: BP2HT, BGHT, cuCollections)  (\S\ref{sec:eval}).
Since its open-source release in October 2022, \sysname has been
integrated into NVIDIA Merlin HugeCTR~\cite{wang2022hugectr},
TFRA~\cite{tfra2020}, and NVIDIA RecSys Examples~\cite{nvidia2024recsys}.
This paper presents the design principles, formal properties, and comprehensive evaluation of \sysname's cache-semantic architecture.

\enlargethispage{2\baselineskip}
\looseness=-1
This paper contributes:
\begin{itemize}[nosep,leftmargin=*]
  \item We analyze GPU embedding cache workloads and identify four challenges (C1--C4) motivating a shift from dictionary to cache semantics (\S\ref{sec:background}).

  \item We design a GPU cache-line-aligned bucket that achieves \emph{definitive per-bucket miss in one fixed memory transaction}: single-bucket confinement collapses each key's entire candidate space into 128~one-byte digests packed into one GPU L1 cache line---a single cache-line load constitutes a complete negative lookup, not merely per-step acceleration within a variable-length probe chain---an alignment infeasible on 64\,B CPU cache lines~(\S\ref{ssec:bucket}).

  \item We introduce an in-line score-driven upsert path that resolves every full-bucket insertion \emph{inside the table}---fusing score comparison, admission control, and Compare-And-Swap commit into a single GPU operation---eliminating the separate eviction data structures, multi-kernel pipelines, and CPU involvement required by all prior eviction-aware designs~\cite{metreveli2011cphash,fan2013memc3,hetherington2015memcachedgpu,fbgemm2019}~(\S\ref{ssec:eviction}).

  \item We introduce score-based dynamic dual-bucket selection, which extends the power-of-two-choices paradigm~\cite{azar1994balanced,mitzenmacher2001power} from load balancing to eviction-quality optimization in GPU hash tables: at full capacity, load-based P2C degenerates to a random coin flip because both buckets are equally full; \sysname restores a meaningful selection criterion by comparing minimum scores across candidate buckets---a decision dimension meaningful only under cache semantics---recovering the ${\sim}34\%$ HBM wasted by birthday-paradox collisions and achieving 99.4\% top-$N$ score retention at $\lambda{=}1.0$~(\S\ref{ssec:dualbucket}).

  \item We design triple-group concurrency with reader/updater/inserter role separation that distinguishes structural from non-structural GPU writes---a split plane absent from all prior role-based designs~\cite{triplett2011rcu,fatourou2018waitfree}---coordinated by a CPU--GPU dual-layer lock---to our knowledge the first hash table concurrency protocol spanning CPU and GPU execution domains---yielding up to $4.80\times$ throughput over R/W locking~(\S\ref{ssec:triplegroup}).

  \item We implement tiered KV separation as a scaling enabler for hybrid HBM+HMEM deployment (\S\ref{ssec:kvsep}), evaluate on H100~NVL against four baselines (\S\ref{sec:eval}), and report open-source integrations in HugeCTR~\cite{wang2022hugectr}, TFRA~\cite{tfra2020}, and NVIDIA RecSys Examples~\cite{nvidia2024recsys} since October~2022.
\end{itemize}

%% file: sections/s2-background.tex
\section{Background and Motivation}
\label{sec:background}

\looseness=-1
We survey embedding storage demands (\S\ref{ssec:embedding}), the GPU hash table design space (\S\ref{ssec:ht-design-space}), dictionary-semantic performance pathology at high load factors (\S\ref{ssec:workload}), and four challenges motivating \sysname's cache-semantic architecture (\S\ref{ssec:challenges}).

\subsection{Embedding Storage in Recommendation Systems}
\label{ssec:embedding}

\begin{figure}[t]
  \centering
  \includegraphics[width=\columnwidth]{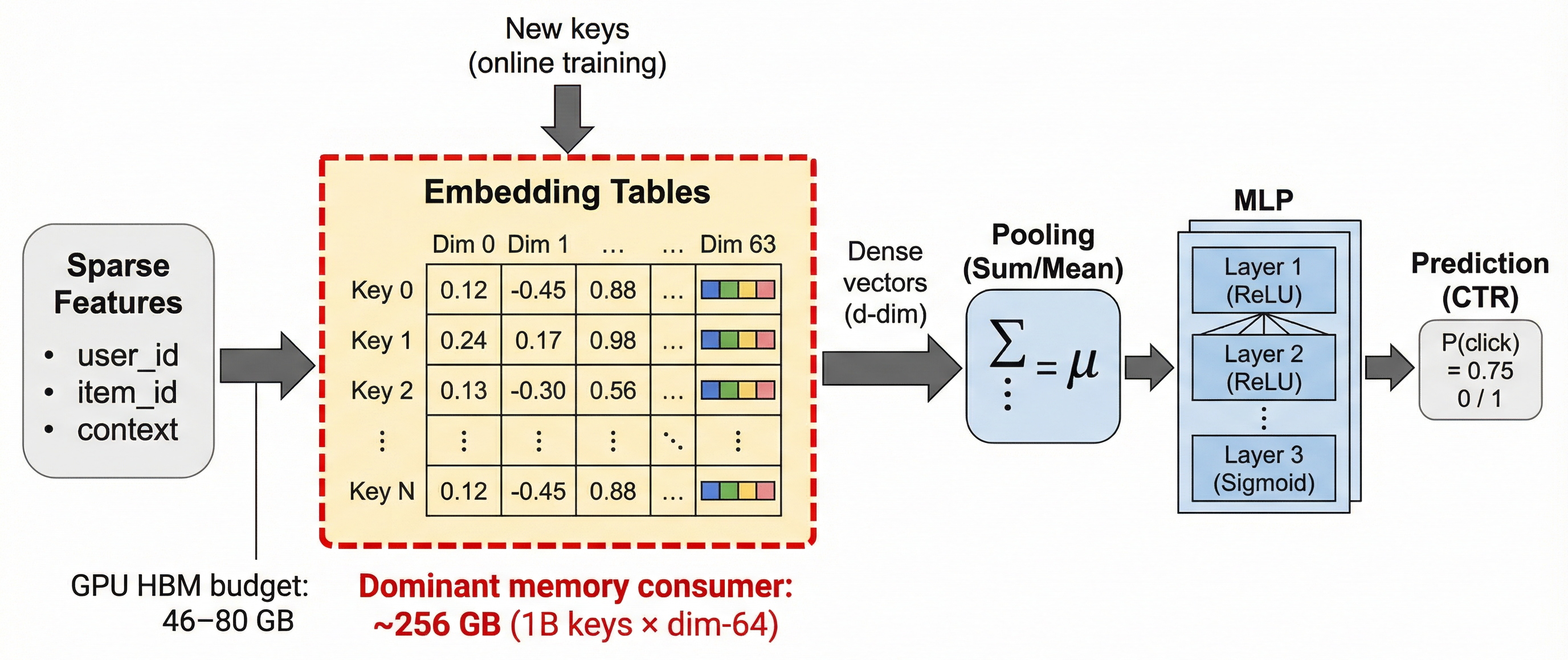}
  \caption{Embedding lookup pipeline in a recommendation model.  Sparse categorical
    features are mapped to dense vectors via embedding tables, which
    constitute the dominant memory consumer of the model.  Online
    training continuously ingests new keys under a hard memory budget.}
  \label{fig:recsys-embedding}
\end{figure}

\looseness=-1
Feature IDs occupy a sparse \texttt{uint64} key space with power-law access~\cite{naumov2019dlrm}, making embedding storage a natural fit for \emph{cache semantics} (Figure~\ref{fig:recsys-embedding}).

\subsection{GPU Hash Table Design Space}
\label{ssec:ht-design-space}

GPU hash tables serve as the indexing substrate for embedding
storage.  Their design involves three interrelated decisions:
collision resolution strategy, probe distance bound, and semantic
commitment (dictionary vs.\ cache).

\looseness=-1
\textbf{Open addressing vs.\ bucketed designs.}
Open-addressing schemes (WarpCore~\cite{junger2020warpcore}, WarpSpeed~\cite{mccoy2025warpspeed}, cuCollections~\cite{nvidia2020cucollections}) probe consecutive slots on collision with unbounded probe distances.
Bucketed designs (Hive~\cite{polak2025hive}, BP2HT~\cite{awad2022better}) group slots into fixed-size buckets, bounding probe distance.
All adopt \emph{dictionary semantics}: every key must be preserved, and a full table triggers rehashing or insertion failure---unacceptable for an embedding cache under continuous ingestion.
Table~\ref{tab:ht-design-space} summarizes the design space; no existing GPU hash table provides built-in eviction or a policy interface.

\begin{table}[t]
  \caption{GPU hash table design space.  All prior GPU hash tables
    adopt dictionary semantics with no built-in eviction.
    \sysname instead adopts a cache-semantic full-capacity contract,
    integrating score-driven admission and eviction directly into the
    insert path.}
  \label{tab:ht-design-space}
  \centering\footnotesize
  \begin{tabular}{@{}llclcc@{}}
    \toprule
    System & Collision & Probe & Sem. & Evict. & LF \\
    \midrule
    WarpCore~\cite{junger2020warpcore}
      & Open addr. & Unbounded & Dict. & No & ${<}1.0$ \\
    WarpSpeed~\cite{mccoy2025warpspeed}
      & Open addr. & Unbounded & Dict. & No & ${<}1.0$ \\
    cuColl.~\cite{nvidia2020cucollections}
      & Open addr. & Unbounded & Dict. & No & ${<}1.0$ \\
    BGHT~\cite{awad2022better}
      & Bucketed & $k{\times}b$ & Dict. & No & ${\sim}.85$ \\
    BP2HT$^\ddagger$~\cite{awad2022better}
      & Bkt.\ P2C & $2{\times}16$ & Dict. & No & ${\sim}.90$ \\
    Hive$^\dagger$~\cite{polak2025hive}
      & Bkt.\ cuckoo & $2{\times}32$ & Dict. & No$^*$ & ${\sim}.95$ \\
    \textbf{\sysname}
      & \textbf{Bucketed} & \textbf{128} & \textbf{Cache}
      & \textbf{Yes} & $\mathbf{1.0}$ \\
    \bottomrule
  \end{tabular}

  \smallskip
  {\footnotesize
  $\dagger$\,Hive: arXiv 2025-10-16 (concurrent/subsequent work; \sysname has been open-source since October~2022).\\
  $^*$\,Hive uses bounded cuckoo relocation + overflow stash
  (dictionary semantics); no score-driven eviction.\\
  $\ddagger$\,The original BP2HT paper~\cite{awad2022better} reports
  $\lambda{\le}0.82$ with 16-slot buckets; our benchmark confirms
  stable throughput through $\lambda{=}0.90$ (Figure~\ref{fig:lf-degradation}).}
\end{table}

\subsection{Workload Analysis: Continuous Online Ingestion}
\label{ssec:workload}

\begin{figure}[t]
  \centering
  \includegraphics[width=\columnwidth]{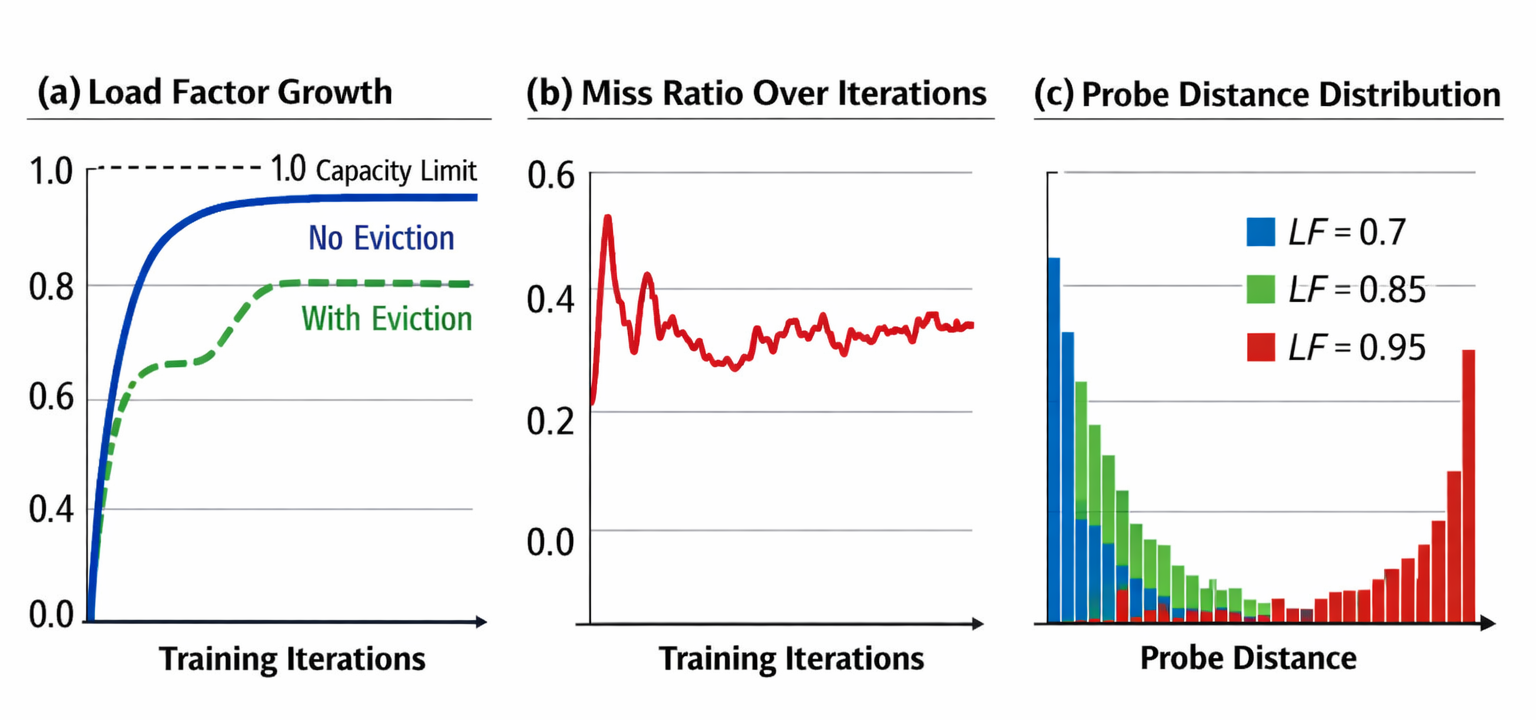}
  \caption{Workload characteristics of continuous online embedding
    ingestion.  (a)~Load factor increases monotonically as new features
    arrive; without eviction, it reaches the capacity ceiling.
    (b)~Miss ratio remains high because new (unseen) feature IDs
    dominate during exploration phases.
    (c)~In open-addressing schemes, probe distance grows
    super-linearly beyond load factor~0.8, causing warp divergence
    on GPUs.}
  \label{fig:workload-analysis}
\end{figure}

Figure~\ref{fig:workload-analysis} summarizes three key workload
characteristics of continuous online ingestion.

\looseness=-1
\textbf{Throughput degradation in dictionary-semantic hash tables.}
Under sustained insertion, probe distances lengthen as $\lambda$ increases, and once full the system must rehash or fail.
As $\lambda$ rises from 0.25 to 1.00, WarpCore \texttt{find} drops 90\%, BGHT 31\%, and cuCollections 100\% (Figure~\ref{fig:lf-degradation}); \sysname remains within ${<}1\%$.

\subsection{Challenges}
\label{ssec:challenges}

The analysis above motivates four design challenges:

\begin{description}[nosep,leftmargin=*,font=\bfseries]
  \item[C1: Fixed-work lookup (\S\ref{ssec:bucket}).]
    Dictionary-semantic hash tables suffer 31--100\% throughput degradation as $\lambda{\to}1.0$ (Figure~\ref{fig:lf-degradation}), yet embedding caches must operate continuously at high~$\lambda$.
    Each operation must therefore complete in a fixed number of memory transactions independent of~$\lambda$.

  \item[C2: In-place full-capacity upsert (\S\ref{ssec:eviction}).]
    Continuously arriving new embeddings require full-bucket inserts to be resolved in place.
    Dictionary-semantic designs either fail the insertion or rehash---blocking the GPU and temporarily doubling memory---neither is acceptable under a hard HBM budget.
    Cache-semantic insertion must instead resolve each full bucket by in-line eviction or admission rejection.

  \item[C3: Retention under bounded associativity (\S\ref{ssec:dualbucket}).]
    Evictions are not equal: evicting a popular embedding forces costly recomputation or a remote fetch.
    Under single-bucket confinement, the birthday paradox wastes ${\sim}34\%$ of capacity; a dual-bucket scheme with score-aware bucket selection is needed to maximize retention of high-value entries.

  \item[C4: Mixed-workload concurrency (\S\ref{ssec:triplegroup}).]
    Production systems issue reads and writes simultaneously from overlapping inference and training kernels~\cite{wang2022hugectr}.
    With up to 270\,K resident GPU threads, coarse-grained R/W locks that serialize non-structural value updates against structural inserts become a first-order bottleneck.
\end{description}

\looseness=-1
Practical embedding tables still often exceed single-GPU HBM capacity.  We therefore treat tiered key-value separation (\S\ref{ssec:kvsep}) as a scaling enabler layered on top of the core cache-semantic contract, rather than as one of the four core challenges above.

\paragraph{Cache-Semantic Hash Tables.}
\begin{definition}[Cache-Semantic Hash Table]
A hash table $H$ is \emph{cache-semantic} if it satisfies:
\textbf{(CS1)}~every full-bucket insert attempt is resolved in place by policy-driven eviction or admission rejection;
\textbf{(CS2)}~no operation triggers rehashing or external capacity management;
\textbf{(CS3)}~lookup cost is bounded independent of the cumulative number of insertions.
\end{definition}
No dictionary-semantic hash table satisfies all three properties: CS1 requires in-table full-capacity resolution, which dictionary semantics forbids.  The remainder of this paper designs and evaluates a GPU hash table library whose normal full-capacity operating contract satisfies CS1--CS3.

%% file: sections/s3-design.tex
\section{System Design}
\label{sec:design}

\looseness=-1
\sysname departs from \emph{dictionary semantics}---where every inserted key
must be preserved and a full table triggers rehashing or insertion
failure---and embraces \emph{cache semantics}, where eviction is a
first-class, policy-driven operation.  Under continuous online ingestion
at load factor~$\lambda{=}1.0$, each full-bucket upsert is resolved in
place by score-driven eviction or admission rejection; there is no
capacity-induced insertion failure, no rehashing on the
critical path, and no external maintenance workflow.
MemcachedGPU~\cite{hetherington2015memcachedgpu} provides cache semantics
at the network-service level with a single fixed Least Recently Used (LRU) policy and CPU on the
eviction path; \sysname instead operates as a library-level GPU hash table
primitive with single-CAS eviction, five pluggable scoring policies, and
zero-cost admission control---all resolved entirely on the GPU.

\subsection{Architecture Overview}
\label{ssec:arch}

\begin{figure}[t]
  \centering
  \includegraphics[width=\columnwidth]{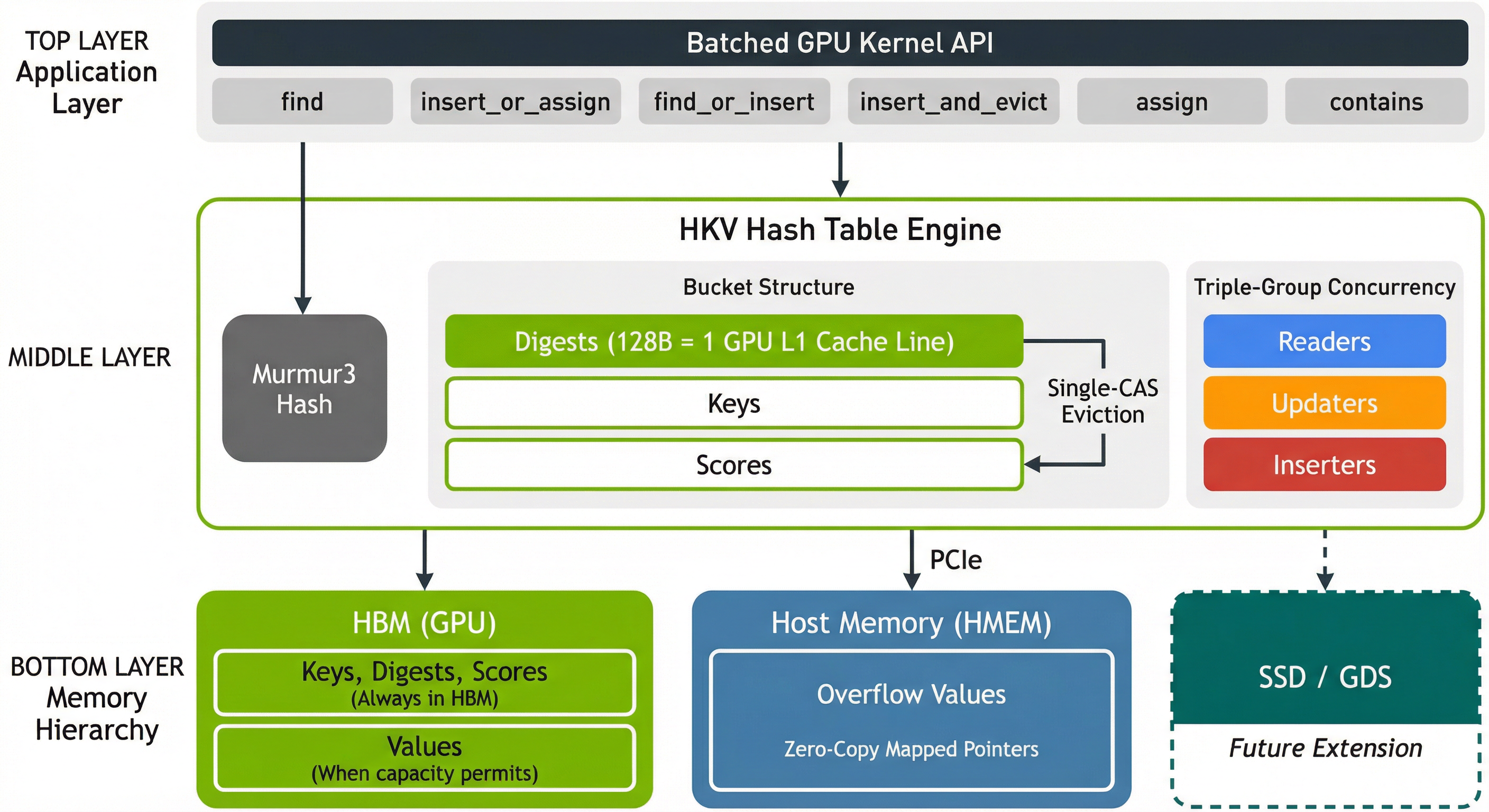}
  \caption{\sysname architecture.  Keys, digests, and scores reside
    in HBM; overflow values are placed in pinned host memory (HMEM)
    via zero-copy mapped pointers.  The SSD/GDS tier (dashed) is an architectural extension point.}
  \label{fig:architecture}
\end{figure}

\looseness=-1
\sysname is a GPU hash table library exposing 17 Standard Template Library (STL)-compatible APIs (\texttt{find}, \texttt{insert\_or\_assign}, \texttt{find\_or\_insert}, \texttt{insert\_and\_evict}, \texttt{assign}, \texttt{contains}, etc.) through batched kernel invocations processing millions of key--value pairs per launch.  As shown in Figure~\ref{fig:architecture}, each request computes Murmur3 hashes, locates target buckets in HBM, and performs digest-accelerated lookups; for \texttt{insert\_or\_assign}, the kernel additionally scans scores when the bucket is full and resolves the upsert in place by admission control plus CAS commit---completing lookup, victim selection, and insertion in a single kernel launch.

\looseness=-1
\textbf{Co-design philosophy.}
The four core mechanisms---single-bucket-confined bucket (\S\ref{ssec:bucket}),
in-line score-driven upsert (\S\ref{ssec:eviction}), dynamic dual-bucket selection
(\S\ref{ssec:dualbucket}), and triple-group concurrency
(\S\ref{ssec:triplegroup})---form a dependency chain whose interactions produce three system-level properties (Table~\ref{tab:emergent-properties}) unattainable by any single component: confinement creates fixed miss work; in-line upsert enables full-capacity resolution; dual-bucket recovers birthday-paradox retention loss; and confining structural modifications to the upsert path enables triple-group concurrency.  Tiered KV separation (\S\ref{ssec:kvsep}) extends the design beyond HBM as a scaling enabler.

\begin{table}[t]
\caption{Core properties of the four co-designed mechanisms in \S\ref{ssec:bucket}--\S\ref{ssec:triplegroup}. Tiered KV separation (\S\ref{ssec:kvsep}) is a scaling enabler on top of this core.}
\label{tab:emergent-properties}
\centering
\footnotesize
\setlength{\tabcolsep}{4pt}
\renewcommand{\arraystretch}{0.95}
\begin{tabular}{@{}lccc@{}}
\toprule
System & \makecell{1-txn\\per-bucket miss} &
         \makecell{In-place\\full-capacity\\resolution} &
         \makecell{Concurrent\\R/U/I} \\
\midrule
SwissTable/F14/BBC       & \texttimes$^a$ & \texttimes & n/a \\
MemC3~\cite{fan2013memc3}    & \texttimes$^b$ & \checkmark$^c$ & n/a \\
MemcachedGPU~\cite{hetherington2015memcachedgpu} & \texttimes & \checkmark$^c$ & \texttimes$^d$ \\
FBGEMM TBE~\cite{fbgemm2019} & \texttimes & \checkmark$^c$ & \texttimes \\
WarpCore/BGHT/BP2HT      & \texttimes & \texttimes & \texttimes \\
\textbf{\sysname}             & \checkmark & \checkmark & \checkmark \\
\bottomrule
\end{tabular}

\smallskip
{\scriptsize
$^a$\,Variable-length probes.~$^b$\,Two-bucket cuckoo miss path.~$^c$\,Fixed-policy in-place resolution, not HKV's GPU upsert contract.~$^d$\,CPU on the critical path.}
\end{table}

\subsection{Single-Bucket-Confined Cache-Line-Aligned Design}
\label{ssec:bucket}

\looseness=-1
Prior GPU hash tables use variable-length probe chains~\cite{junger2020warpcore,awad2022better,polak2025hive} or multi-bucket designs~\cite{fan2013memc3,hegeman2024compact,herlihy2008hopscotch}; \sysname covers all 128 candidates in a single cache-line load (Table~\ref{tab:miss-cost}).

\looseness=-1
The key design principle of \sysname's bucket structure is to
\emph{collapse the entire candidate space of every key into a single,
GPU L1 cache-line-aligned unit}, so that a miss within a bucket can be
determined definitively in one fixed memory transaction.
(Under dual-bucket mode (\S\ref{ssec:dualbucket}), two fixed-length
transactions cover both candidate buckets.)

\looseness=-1
\textbf{Single-bucket confinement.}
\sysname maps each key by a GPU-optimized hash derived from Murmur3 to exactly one
bucket of 128~slots---no secondary hash, no overflow chain, no
cuckoo relocation.  The bucket is the key's \emph{entire}
candidate space.
Unlike bounded-probe designs that still require secondary
probes~\cite{herlihy2008hopscotch,panigrahy2005efficient,breslow2016horton,awad2022better,polak2025hive},
single-bucket confinement guarantees a definitive miss after scanning
one bucket.
\looseness=-1
\textbf{Cache semantics as structural enabler.}
Single-bucket confinement is unachievable under dictionary
semantics at high load factors: a full bucket with no overflow
chain causes insertion failure.  Cache semantics addresses
this---score-driven eviction deterministically replaces the
minimum-score entry, eliminating overflow chains, rehashing, and
capacity-induced failures.  This semantic shift is what makes
confinement viable at $\lambda{=}1.0$ and, in turn, enables the
definitive miss property described below.
Adas et al.~\cite{friedman2021limited} show that bounded associativity with eviction maintains $O(1)$ worst-case lookup.
On GPUs, single-hash lookup eliminates three architecture-specific
costs:
(1)~warp divergence from branching on ``which bucket holds the
key?'';
(2)~the computation cost of a second independent hash function
across millions of keys per kernel launch; and
(3)~non-coalesced memory access from probing two non-adjacent
buckets, which doubles L1/L2 cache pressure per lookup.

\looseness=-1
\textbf{Cache-line-aligned digest array.}
\sysname couples single-bucket confinement with inline fingerprinting
for the GPU memory hierarchy: each slot stores an 8-bit
digest (bits 32--39 of the GPU-optimized Murmur3 variant).
\sysname separates digests into a contiguous array purpose-built for
the GPU's 128\,B L1 cache line: 128~one-byte digests $=$ one cache-line
load, enabling 128~candidates in a single memory transaction.
CPU hash tables---SwissTable~\cite{swisstable2017} (7-bit H2 per group, SSE scan) and F14~\cite{f14folly2019} (7-bit tag, SIMD chunk filtering)---embed tags within fixed-size groups of at most 16~entries per 64\,B CPU cache line; BBC~\cite{bother2023bbc} places a contiguous 16-fingerprint sub-array at each bucket head for SIMD scanning, but its buckets span multiple cache lines and still require overflow probing at high load factors. In contrast,
MemcachedGPU~\cite{hetherington2015memcachedgpu} embeds 8-bit hashes in ${\ge}$20\,B per-entry headers, requiring up to 16~candidate slots across multiple cache-line loads per miss.
\sysname achieves $8\times$ the candidate coverage per transaction versus both CPU and prior GPU designs.
A lookup performs 32 \texttt{\_\_vcmpeq4} comparisons; only digest-matching slots trigger full key comparison (false-positive rate $1/256$, ${\sim}0.5$ unnecessary key reads per miss).

\begin{figure}[t]
  \centering
  \includegraphics[width=0.75\columnwidth]{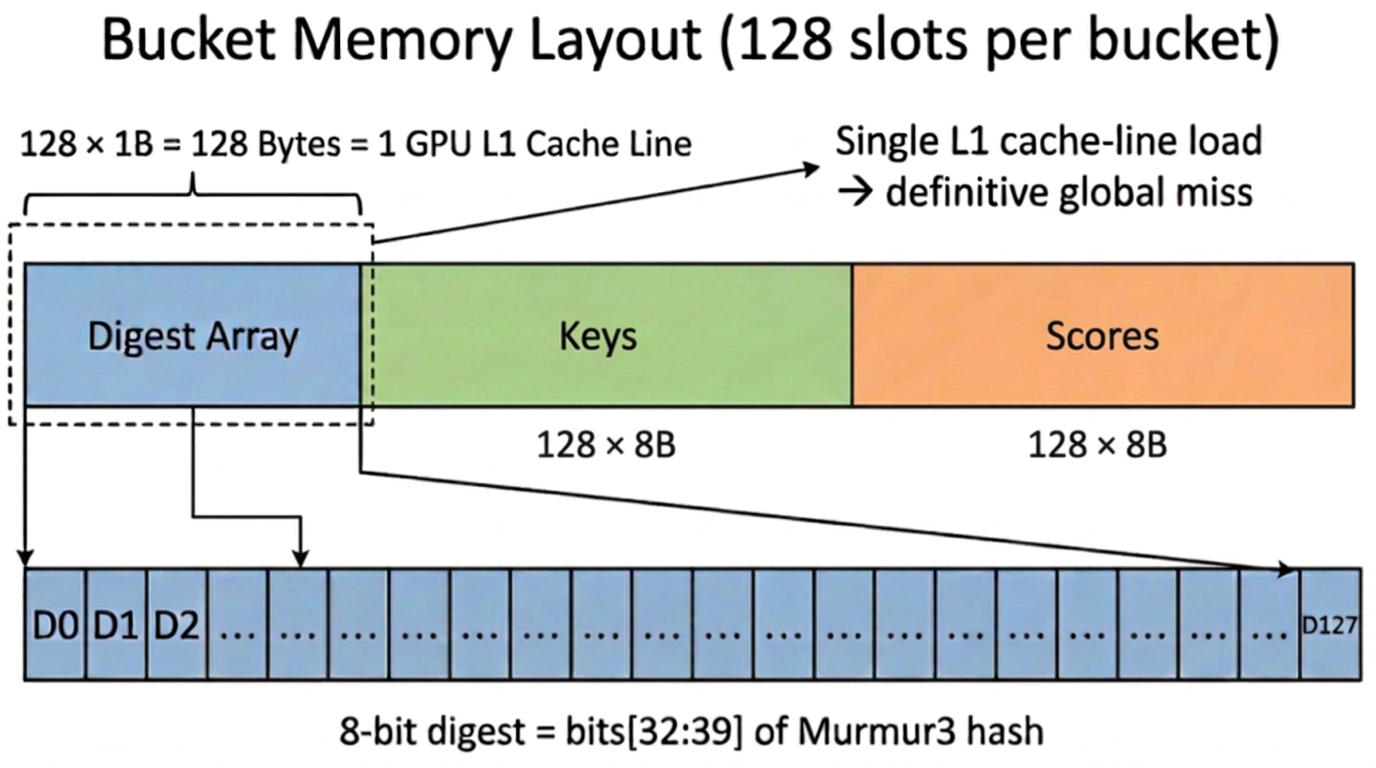}
  \caption{Memory layout of a single \sysname bucket (128~slots).
    The digest array occupies exactly one GPU L1 cache line
    (128\,B), enabling a complete per-bucket negative lookup in a
    single cache-line load.
    Values are addressed by bucket and slot index
    (position-based addressing, \S\ref{ssec:kvsep}); no per-entry
    pointer is stored.
}
  \label{fig:bucket-layout}
\end{figure}

\looseness=-1
\textbf{Definitive per-bucket miss in one memory transaction.}
Three co-designed decisions collapse the miss-path cost to one cache-line load per bucket (Figure~\ref{fig:bucket-layout}): (1)~contiguous digest separation packs 128 digests into one 128-byte GPU L1 cache line; (2)~single-bucket confinement makes these 128~slots the key's entire candidate space; (3)~one load therefore covers all candidates.  In single-bucket mode this yields a table-level definitive miss because the bucket is the entire candidate space; dual-bucket mode (\S\ref{ssec:dualbucket}) extends to two buckets (two fixed-length transactions) while improving retention quality.

Algorithm~\ref{alg:find} summarizes the digest-accelerated
find path for a single bucket.

\begin{proposition}[Definitive Per-Bucket Miss]
\label{prop:definitive-miss}
In single-bucket mode with $S{=}128$ slots per bucket, for any key $k$ not present in $T$, \textsc{Find}$(k)$ (Algorithm~\ref{alg:find}) returns \textsc{NotFound} after examining exactly one bucket, performing $S$ digest comparisons and at most $S \cdot 2^{-8} = 0.5$ expected full-key comparisons, using a single 128-byte memory transaction.
\end{proposition}

\begin{algorithm}[t]
\small
\caption{Digest-Accelerated Find}\label{alg:find}
\begin{algorithmic}[1]
\Require Key $k$, hash table $T$ with $B$ buckets of $S{=}128$ slots each
\Ensure \textsc{Found}$(v)$ or \textsc{NotFound}
\State $b \gets \textsc{Hash}(k) \bmod B$
\Comment{Map to bucket}
\State $d \gets \textsc{Hash}(k)[31{:}24]$
\Comment{8-bit digest}
\State $D[0..S{-}1] \gets T.digests[b]$
\Comment{One 128\,B cache-line load}
\For{$i \gets 0$ \textbf{to} $S/4-1$}
    \Comment{SIMD 4-way compare}
    \State $m \gets \textsc{\_\_vcmpeq4}(D[4i{:}4i{+}3],\; d \times \texttt{0x01010101})$
    \For{each set byte $j$ in $m$}
        \If{$T.keys[b][4i{+}j] = k$}
            \State \Return \textsc{Found}$(T.values[b][4i{+}j])$
        \EndIf
    \EndFor
\EndFor
\State \Return \textsc{NotFound}
\Comment{Definitive miss---all $S$ slots checked}
\end{algorithmic}
\end{algorithm}

\looseness=-1
Table~\ref{tab:miss-cost} quantifies this distinction.  All prior
designs incur miss costs that grow with load factor or require
rehashing at full capacity; \sysname delivers constant-cost misses
at every load factor including $\lambda{=}1.0$.

\begin{table}[t]
  \caption{Miss-path I/O cost (memory loads per negative lookup) at
    representative load factors.  ``---'' indicates the load factor
    is unreachable (rehash required).  Prior-system rows are schematic
    structural counts derived from published probe organizations rather
    than same-platform benchmark measurements.  \sysname is the only
    design with load-factor-independent miss cost.}
  \label{tab:miss-cost}
  \centering
  \footnotesize
  \setlength{\tabcolsep}{4pt}
  \begin{tabular}{@{}lccc@{}}
    \toprule
    Design & $\lambda{=}0.50$ & $\lambda{=}0.875$ & $\lambda{=}1.0$ \\
    \midrule
    CPU (SwissTable, F14, BBC)~\cite{swisstable2017,f14folly2019,bother2023bbc}
      & ${\sim}$2 loads & ${\sim}$8 loads & --- \\
    MemC3~\cite{fan2013memc3}
      & 2 bkt loads & 2 bkt loads & --- \\
    \midrule
    \multicolumn{4}{l}{\emph{GPU hash tables}} \\
    WarpCore~\cite{junger2020warpcore}
      & ${\sim}2$ probes & ${\sim}8$ probes & --- \\
    BGHT~\cite{awad2022better}
      & $2{\times}16$ slots & $2{\times}16$ slots & --- \\
    BP2HT~\cite{awad2022better}
      & $2{\times}16$ slots & $2{\times}16$ slots & --- \\
    Hive~\cite{polak2025hive}
      & $2{\times}32$ slots & $2{\times}32$ slots & --- \\
    \midrule
    \textbf{\sysname (single)}
      & \textbf{1 CL load} & \textbf{1 CL load} & \textbf{1 CL load} \\
    \textbf{\sysname (dual)}
      & \textbf{2 CL loads} & \textbf{2 CL loads} & \textbf{2 CL loads} \\
    \bottomrule
  \end{tabular}
\end{table}

Single-bucket confinement guarantees constant-cost misses; the next subsection addresses what happens when a confined bucket fills.

\subsection{Score-Driven Built-in Eviction}
\label{ssec:eviction}

\looseness=-1
Under cache semantics, a full bucket is not an error---it is the
\emph{steady state}.  To our knowledge, \sysname is the first GPU hash table to fuse eviction directly into
the insert path so that no external eviction workflow (export,
sort, compact) is ever needed.

\looseness=-1
\textbf{Architectural distinction from prior eviction-aware designs.}
In all prior hash tables with built-in
eviction---CPHash~\cite{metreveli2011cphash} (LRU list,
16\,B/entry pointer overhead), MemC3's CLOCK~\cite{fan2013memc3}
(per-entry reference bits with external clock-hand sweep), CacheLib~\cite{berg2020cachelib} (intrusive
hook lists, ${\sim}$24\,B/entry), and FBGEMM TBE~\cite{fbgemm2019} (multi-kernel eviction
pipeline)---eviction metadata forms a \emph{second data structure} with
its own memory layout, traversal logic, and synchronization domain.
\sysname eliminates this second structure entirely: the score array
\emph{is} the eviction metadata, the bucket \emph{is} the eviction
scope, and the CAS \emph{is} the eviction lock---a unification absent from all prior GPU hash tables.
HashPipe~\cite{sivaraman2017hashpipe} embeds count-based eviction into a pipeline of single-entry hash stages on a switch ASIC; \sysname shares the principle of in-path eviction but operates on 128-slot GPU buckets with pluggable scoring and general key--value semantics.

\looseness=-1
The contrast with FBGEMM TBE---the closest industrial system---is
illustrative:
(i)~FBGEMM requires a \emph{multi-kernel} pipeline (separate kernels
for cache-miss resolution, per-set victim selection, and replacement) vs.\ \sysname's single in-line
upsert path with CAS commit;
(ii)~FBGEMM supports 2~fixed eviction policies (LRU, Least Frequently Used (LFU)) vs.\
\sysname's built-in score variants plus a caller-supplied custom-score
path;
(iii)~FBGEMM's associativity is 8--32 ways (version-dependent) vs.\ \sysname's 128-slot
GPU L1-aligned bucket with digest-accelerated victim search.
A further departure is \emph{admission control}: when the incoming
key's score is lower than the bucket minimum, the insertion is
refused---the table retains the higher-value entry, preserving cache
quality under adversarial patterns.  Unlike TinyLFU's~\cite{einziger2017tinylfu} frequency-sketch-based admission gate, \sysname's admission check is embedded in the bucket-local upsert path with zero additional data structures.

\looseness=-1
\textbf{Scoring policies.}
A compile-time \texttt{ScoreFunctor} abstraction realizes LRU, LFU, epoch-aware, and custom scoring through the same in-line upsert mechanism---without a second eviction data structure (\S\ref{sec:eval}).

\looseness=-1
\textbf{In-line bucket-local admission and commit.}
When a bucket is full, the kernel scans all
128~scores, identifies the minimum-score slot, and atomically
replaces it via CAS on the key field.
The scan uses asynchronous prefetch with double-buffered shared
memory to hide HBM latency (\S\ref{sec:impl-kernel}).
Algorithm~\ref{alg:upsert} formalizes the complete upsert path.

\begin{algorithm}[t]
\small
\caption{Upsert with Bucket-Local Admission and Eviction.  Free slots trigger
  direct insertion (line~8); full buckets undergo eviction with
  admission control (lines~10--13); CAS serves as lock and commit
  (line~14).}\label{alg:upsert}
\begin{algorithmic}[1]
\Require Key $k$, value $v$, score $s$, hash table $T$
\Ensure \textsc{Inserted}, \textsc{Updated}, \textsc{Rejected}, or \textsc{Evicted}$(k_{\text{old}})$
\State Run Algorithm~\ref{alg:find} for $k$ in bucket $b$
\If{found at slot $j$}
    \State $T.values[b][j] \gets v$;\; $T.scores[b][j] \gets s$
    \State \Return \textsc{Updated}
\EndIf
\State $e \gets$ first empty slot in bucket $b$
\If{$e \neq \textsc{Nil}$}
    \State Write $(k, v, s, d)$ to slot $e$
    \State \Return \textsc{Inserted}
\EndIf
\Comment{Bucket full---eviction path}
\State $m \gets \arg\min_{i} T.scores[b][i]$
\If{$s < T.scores[b][m]$}
    \State \Return \textsc{Rejected}
    \Comment{Admission control}
\EndIf
\If{$\textsc{CAS}(T.keys[b][m],\; k_{\text{old}},\; \textsc{Locked}) = k_{\text{old}}$}
    \State Write $(k, v, s, d)$ to slot $m$
    \State \Return \textsc{Evicted}$(k_{\text{old}})$
\Else
    \State \textbf{goto} line 11
    \Comment{Retry on CAS failure}
\EndIf
\end{algorithmic}
\end{algorithm}

\begin{proposition}[Liveness]\label{prop:eviction-progress}
If bucket $b$ is full and the inserting thread holds the exclusive lock,
then Algorithm~\ref{alg:upsert} terminates in $O(B)$ time, where
$B{=}128$ is the bucket capacity.  If additionally $s \geq s_{\min}(b)$,
the new entry replaces the minimum-score entry via a single CAS.
Under the triple-group protocol (\S\ref{ssec:triplegroup}), at most one inserter kernel executes concurrently per table.  Because each warp cooperatively processes exactly one key and its corresponding bucket, no two warps operate on the same bucket simultaneously; CAS contention is therefore confined to threads within a single warp, bounding retries to at most $W{-}1{=}31$ per thread (warp width $W{=}32$).
\end{proposition}

\subsection{Score-Based Dynamic Dual-Bucket Selection}
\label{ssec:dualbucket}

\looseness=-1
Single-bucket confinement (\S\ref{ssec:bucket}) introduces a
fundamental memory-utilization problem: with 128-slot buckets,
the birthday paradox triggers the first eviction at
$\lambda{\approx}0.66$~\cite{walzer2023load}, leaving \emph{${\sim}34\%$ of allocated
HBM unused} before the eviction path even activates.
On GPUs where HBM capacity directly constrains model size, this
waste is unacceptable.
\sysname addresses this with a dynamic dual-bucket mode that
extends the power-of-two-choices
paradigm~\cite{seznec1993skewed,azar1994balanced,mitzenmacher2001power}
from load-based to \emph{score-based} selection.
Each key maps to two candidate buckets via independent hashes.
Unlike BP2HT~\cite{awad2022better} and other load-based two-choice designs, \sysname introduces a two-phase adaptive policy that shifts the selection criterion from load balancing to \emph{eviction quality}---selecting the bucket whose minimum score yields the best eviction decision.  The two phases are:

\looseness=-1
\textbf{Phase~D1 (warm-up): memory utilization.}
While at least one candidate bucket has a free slot, the key
is inserted into the less-loaded bucket---classic load-balanced
two-choice placement that delays the first eviction from
$\lambda{\approx}0.66$ to $\lambda{>}0.97$
(\S\ref{ssec:exp-bucket}), recovering nearly all wasted HBM.

\looseness=-1
\textbf{Phase~D2 (steady state): score-based selection.}
At $\lambda{\approx}1.0$ both buckets are full, so load-based P2C degenerates to a coin flip.
\sysname restores the two-choice benefit by shifting the decision dimension from load to \emph{score}: the kernel evicts in the bucket with the lower minimum score (Algorithm~\ref{alg:dualbucket}).
Under sustained Zipfian ingestion, dual-bucket mode yields 99.4\% top-$N$ score retention versus 95.4\% for single-bucket---a 4.05\,pp improvement (\S\ref{ssec:exp-bucket}).

\begin{figure}[t!]
  \centering
  \includegraphics[width=0.95\columnwidth]{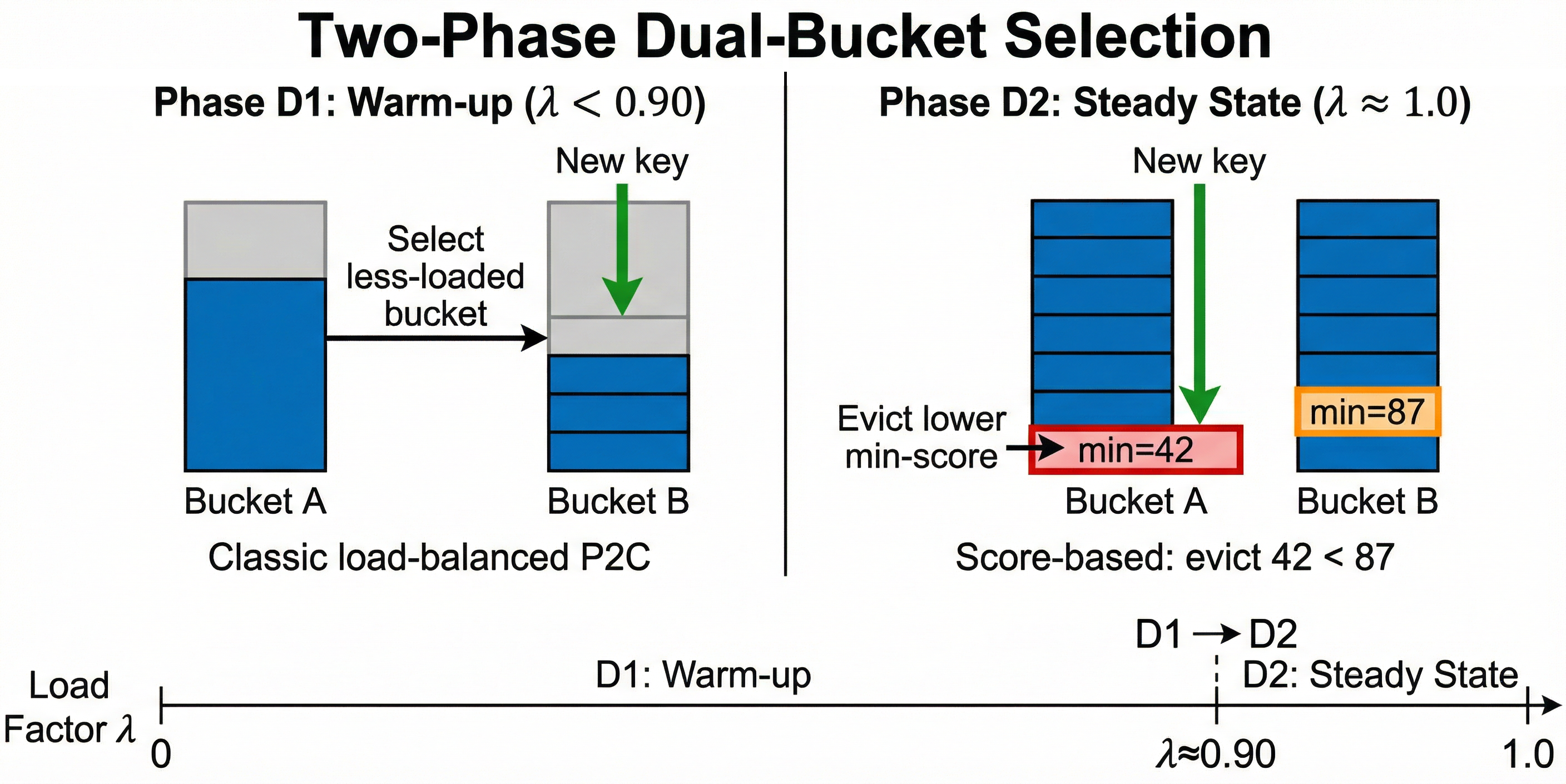}
  \caption{Two-phase dual-bucket selection.  Phase~D1 (left) inserts
    into the less-loaded bucket for memory utilization; Phase~D2
    (right, $\lambda{\approx}1.0$) evicts in the bucket with the
    lower minimum score, improving eviction correctness.}
  \label{fig:dual-bucket}
\end{figure}

\looseness=-1
\textbf{Lookup cost analysis.}
Dual-bucket mode issues two cache-line transactions instead of one on the miss path---still $O(1)$ and load-factor-independent.  Despite the extra transaction, dual-bucket achieves comparable or higher throughput (\S\ref{ssec:exp-bucket}) by avoiding premature eviction overhead starting at $\lambda{\approx}0.66$.
Figure~\ref{fig:dual-bucket} illustrates the two-phase policy.

\begin{algorithm}[t]
\small
\caption{Score-Based Dual-Bucket Upsert}\label{alg:dualbucket}
\begin{algorithmic}[1]
\Require key $k$, value $v$, score $s$
\State $b_1 \gets h_1(k) \bmod B$;\; $b_2 \gets h_2(k) \bmod B$ \Comment{two candidate buckets}
\If{$\text{occupancy}(b_1) < 128$ \textbf{or} $\text{occupancy}(b_2) < 128$} \Comment{Phase D1}
  \State \textbf{insert} $(k,v,s)$ into less-occupied bucket
\Else \Comment{Phase D2: score-based ($\lambda{=}1.0$)}
  \State $m_1 \gets \min\text{-score}(b_1)$;\; $m_2 \gets \min\text{-score}(b_2)$
  \State $b^* \gets \arg\min(m_1, m_2)$ \Comment{bucket with lower min-score}
  \If{$s > \min(m_1, m_2)$} \Comment{admission control}
    \State \textbf{call} \textsc{Upsert-With-Eviction}($b^*$, $k$, $v$, $s$) \Comment{Alg.~\ref{alg:upsert}}
  \Else
    \State \textbf{reject} insertion \Comment{incoming score too low}
  \EndIf
\EndIf
\end{algorithmic}
\end{algorithm}

\begin{proposition}[Score-Based Selection Advantage]\label{prop:dualbucket}
Under i.i.d.\ scores drawn from CDF $F$, selecting the bucket with the lower minimum score reduces the expected evicted score from $\mathbb{E}[X_{(1:n)}]$ to $\mathbb{E}[\min(X_{(1:n)}, Y_{(1:n)})]$.  For uniform $F$ on $[0,1]$ with bucket size $n{=}128$, this halves the expected evicted score: $\mathbb{E}[\min] = 1/(2n{+}1) \approx 0.0039$ vs.\ $1/(n{+}1) \approx 0.0078$; see standard order-statistics references such as David and Nagaraja~\cite{david2003order}.
\end{proposition}
\noindent Under production Zipfian workloads, scores are not i.i.d.\ uniform; the empirical improvement (Table~\ref{tab:single-vs-dual}) exceeds the i.i.d.\ bound because high-value entries cluster in the hot bucket.

With bucket design, eviction, and load balancing in place, the
remaining challenge is safe concurrent access from both GPU and CPU.

\subsection{Triple-Group Concurrency Control}
\label{ssec:triplegroup}

\looseness=-1
\sysname introduces a three-group concurrency protocol that
separates operations by whether they modify bucket
\emph{structure}---a split plane absent from all prior
role-based hash table designs.
No prior role-based design---whether ``read vs.\ write'' (Read-Copy-Update (RCU)~\cite{triplett2011rcu}) or ``mutation vs.\ resize'' (Fatourou et al.~\cite{fatourou2018waitfree})---distinguishes \emph{structural} from \emph{non-structural} writes, and none spans CPU--GPU execution domains (\S\ref{sec:related}).
\sysname's triple-group scheme isolates \emph{readers}, \emph{updaters}, and \emph{inserters} into distinct GPU kernels, adding a CPU--GPU dual-layer lock for host/device coordination.
The three groups are:

\begin{itemize}[nosep,leftmargin=*]
  \item \textbf{Readers} (\texttt{find}, \texttt{contains}, \texttt{size}):
    multiple readers execute in parallel; each reader kernel assumes
    stable bucket structure and requires no CAS.
  \item \textbf{Updaters} (\texttt{assign}, \texttt{assign\_scores}):
    multiple updaters execute in parallel; they modify existing entries'
    values or scores in place but never alter bucket structure (no slot
    allocation, no digest update, no eviction).
  \item \textbf{Inserters} (\texttt{insert\_or\_assign}, \texttt{find\_or\_insert},
    \texttt{erase}): exclusive access; inserters allocate slots, update
    digests, and perform evictions---structural modifications that are
    incompatible with concurrent reads or updates.
\end{itemize}

\looseness=-1
\noindent
The structural/non-structural distinction is enabled by score-driven
eviction (\S\ref{ssec:eviction}): because eviction, slot allocation,
and digest updates are embedded in the insert path, all structural
modifications are confined to the inserter role.
Single-bucket confinement (Section~\ref{ssec:bucket}) ensures that an insert never triggers cascading relocations across multiple buckets; this containment property makes table-level reader/updater/inserter role isolation sufficient for correctness without per-bucket fine-grained locking.
Fusing eviction into the insert path is likewise essential: if eviction were a separate background process (as in MemC3's CLOCK scan or CacheLib's \texttt{AccessContainer}), it would create structural modifications outside the inserter role, breaking the three-group invariant that confines all structural changes to a single kernel type.
Table~\ref{tab:concurrency} summarizes the resulting compatibility
matrix.

\begin{table}[t]
\centering
\caption{Triple-group concurrency compatibility. \checkmark = concurrent execution permitted; \texttimes = mutually exclusive.}
\label{tab:concurrency}
\begin{tabular}{lccc}
\toprule
 & Reader & Updater & Inserter \\
\midrule
Reader   & \checkmark & \texttimes & \texttimes \\
Updater  & \texttimes & \checkmark & \texttimes \\
Inserter & \texttimes & \texttimes & \texttimes \\
\bottomrule
\end{tabular}
\end{table}

\looseness=-1
\textbf{Concurrency semantics.}
The compatibility matrix (Table~\ref{tab:concurrency}) follows Gray's intent-lock hierarchy~\cite{gray1976granularity}.  The key benefit over a reader/writer lock is that \emph{multiple updaters execute concurrently}: under R/W locking, all value updates serialize as writes, whereas triple-group allows parallel updater batches, yielding up to $4.80\times$ throughput (\S\ref{ssec:exp-ablation}).

Readers and updaters are mutually exclusive because GPU value updates span up to 256\,bytes and cannot be observed atomically without per-entry locking; table-level serialization avoids this overhead.

The acquire/release protocol uses atomic group counters; details appear in \S\ref{sec:impl-lock}.

\begin{proposition}[Reader Safety]
\label{prop:reader-safety}
Under the triple-group protocol,
a reader kernel never observes a bucket in which a key has been evicted
but its replacement has not yet been written.
This follows directly from the compatibility matrix (Table~\ref{tab:concurrency}): no reader and inserter kernel execute concurrently, so the reader cannot observe any intermediate eviction state.
\end{proposition}

\looseness=-1
\textbf{CPU--GPU dual-layer lock.}
GPU kernels cannot directly observe host-side atomic variables, and host
threads cannot control the execution order of warps within a running kernel.
This cross-domain gap necessitates an explicit two-layer synchronization
mechanism that bridges the host/device boundary.
The host layer announces role-group transitions via \texttt{std::atomic}
counters that track active readers, updaters, and inserters;
a lightweight synchronization kernel propagates the committed role state
to device memory, ensuring all GPU warps observe the new role assignment
before the next group launches.
This avoids GPU-side global barriers and device-wide atomic contention
while guaranteeing that incompatible role groups never execute concurrently.
Implementation details appear in Section~\ref{sec:impl-lock}.
With the four core mechanisms in place, one capacity constraint remains: embedding tables that exceed single-GPU HBM.  The next subsection extends the design beyond HBM while keeping key-side processing on GPU.

\subsection{Tiered Key-Value Separation}
\label{ssec:kvsep}

\looseness=-1
Embedding tables often exceed single-GPU HBM capacity.  As a scaling enabler, \sysname implements
\emph{position-based} tiered KV separation:
keys, digests, and scores always reside in HBM; values overflow to
pinned host memory (HMEM) via zero-copy mapped pointers, keeping the
CPU off the critical path.
Because each key maps to exactly one bucket, a value's address is
computed arithmetically from bucket and slot indices---no per-entry
pointer is stored, eliminating 8\,B/entry overhead
(1\,GB at 128\,M capacity).
The architectural consequence is that key-side throughput
(\texttt{find*}, \texttt{contains}) is largely independent of value
placement: on H100~NVL, \texttt{find*} retains 96.0\% of pure-HBM
throughput in hybrid mode.
The architecture accommodates additional storage tiers (e.g., SSD via GPUDirect Storage) below HMEM; the present work focuses on the HBM+HMEM path that covers the dominant deployment scenario.
Implementation details appear in \S\ref{sec:impl-storage}.

Section~\ref{sec:eval} validates these claims experimentally.

%% file: sections/s4-implementation.tex

\section{Implementation}
\label{sec:implementation}

\looseness=-1
This section describes the engineering choices that map \sysname's design (\S\ref{sec:design}) onto GPU hardware.
We cover the public API surface (\S\ref{sec:impl-api}), the tiered storage engine that realizes key-value separation (\S\ref{sec:impl-storage}), GPU kernel engineering (\S\ref{sec:impl-kernel}), and the physical implementation of the triple-group lock (\S\ref{sec:impl-lock}).

\subsection{API Design and Usability}
\label{sec:impl-api}

\looseness=-1
\sysname exposes an STL-style interface centered on a templated
\texttt{Hash\-Table<K,\allowbreak{}V,\allowbreak{}S>} class,
where \texttt{K}~is the key type (typically
\texttt{uint64\_t}), \texttt{V}~the value element type
(e.g., \texttt{float}), and \texttt{S}~the score type
(\texttt{uint64\_t}).

\looseness=-1
\sysname retains familiar map operations
(\texttt{find}, \texttt{erase},
\texttt{insert\_\allowbreak{}or\_\allowbreak{}assign})
and adds cache-specific primitives:
\texttt{insert\_\allowbreak{}and\_\allowbreak{}evict} returns evicted
entries in a single kernel launch;
\texttt{find\_\allowbreak{}or\_\allowbreak{}insert} combines lookup
with insertion for cold-start; and
\texttt{export\_\allowbreak{}batch\_\allowbreak{}if} streams entries
matching a predicate to the host for checkpointing.
Each call acquires the corresponding triple-group lock
(\S\ref{ssec:triplegroup}), launches CUDA kernels, and releases
the lock upon completion.

\subsection{Tiered Storage Engine}
\label{sec:impl-storage}

\looseness=-1
The key-value separation described in \S\ref{ssec:kvsep} is
realized through a slice-based memory allocator.
At initialization, \sysname computes a user-configurable \emph{HBM watermark} that determines the maximum HBM budget for value storage.
Keys, digests, and scores are always placed in HBM; value slices
spill to pinned host memory (HMEM) via CUDA mapped allocation when the watermark is
exhausted.  Pinned pages are mapped into the GPU address space,
so kernels access host-resident values through the same pointer
interface---\emph{no CPU thread is on the data path}.

\looseness=-1
Value storage is organized into coarse-grained \emph{slices}
(256\,KB--16\,GB depending on table capacity) to amortize
pinned-memory allocation overhead.
Within each slice, values are laid out contiguously by bucket;
a value's address is derived from its bucket and slot
index---no per-entry pointer is stored.

\subsection{Adaptive Kernel Selection}
\label{sec:impl-kernel}

\looseness=-1
Maximizing key-level parallelism (one thread per key) yields the highest probe throughput, but large-value copies and high-$\lambda$ eviction scans each demand different thread cooperation.  \sysname provides three upsert kernel variants and a runtime selector based on load factor and value size.

\looseness=-1
\textbf{TLPv1: SIMD digest probing.}
Each thread handles one key independently.
The 128-slot digest array is loaded as 32 vectorized 4-byte words, each compared via \texttt{\_\_vcmpeq4} (byte-parallel SIMD)---covering all 128~slots in 32~instructions with expected ${\sim}0.5$ false-positive key comparisons per miss ($1/256$ per slot).
Value writes use streaming stores (\texttt{\_\_stcs}) to bypass L1.
Selected for small values (${\leq}32$\,B) at $\lambda \leq 0.98$.

\looseness=-1
\textbf{TLPv2: two-phase parallelism switching.}
Probing is identical to TLPv1, but after the probe phase threads regroup into cooperative groups of 8--16 that collectively copy each member's value through double-buffered shared memory---switching from key-level to bandwidth-level parallelism mid-kernel.
The shared-memory buffers are reused between eviction-scan and value-copy phases.
Selected for medium values ($32$--$512$\,B) at $\lambda \leq 0.95$.

\looseness=-1
\textbf{Pipeline: warp-cooperative 4-stage overlap.}
A full 32-thread warp processes keys sequentially, overlapping four stages across consecutive keys:
(1)~async prefetch of 128~digests via \texttt{\_\_pipeline\_memcpy\_async};
(2)~warp-wide SIMD digest comparison;
(3)~cooperative score reduction (\texttt{cg::reduce}) with CAS commit;
(4)~warp-cooperative value writeback.
Three-deep pipeline commits hide HBM latency.
Selected at $\lambda > 0.95$, where frequent eviction scans make latency hiding critical.

\looseness=-1
\textbf{Runtime selection.}
Value size determines the cooperation strategy (TLPv1 vs.\ TLPv2); load factor triggers transition to Pipeline at $\lambda{\approx}0.95$, where latency hiding outweighs per-key serialization cost.
Lookup kernels follow an analogous structure but omit locking via the triple-group guarantee (\S\ref{ssec:triplegroup}).

\subsection{Triple-Group Lock and Key-Level CAS}
\label{sec:impl-lock}

\looseness=-1
The triple-group lock is implemented via three atomic counters on the host and mirrored device-side atomic variables; a lightweight single-thread propagation kernel synchronizes the two layers, reusing a thread-local cached CUDA stream.
The two layers synchronize via a store--launch--fence handshake: the host writes its atomic counter, launches the synchronization kernel, and the kernel's completion constitutes the cross-domain fence.
Fine-grained slot-level mutual exclusion is achieved by atomically replacing a key with a sentinel lock value via CAS with acquire semantics, restoring the actual key after the value write completes.

%% file: sections/s5-evaluation.tex
\section{Evaluation}
\label{sec:eval}

\looseness=-1
We evaluate \sysname through four experiments: load-factor sensitivity against four baselines (\S\ref{ssec:exp-lf}), end-to-end throughput including hybrid HBM+HMEM mode (\S\ref{ssec:exp-e2e}), component ablation covering digest filtering, eviction overhead, cache quality, admission control, and concurrency (\S\ref{ssec:exp-ablation}), and single- vs.\ dual-bucket analysis (\S\ref{ssec:exp-bucket}).

\subsection{Experimental Setup}
\label{ssec:exp-setup}

\textbf{Hardware.}
All experiments run on a single NVIDIA H100 NVL GPU (94\,GB HBM3, Hopper,
Compute Capability~9.0) connected via Peripheral Component Interconnect Express (PCIe)~Gen5 $\times$16, installed in
a server with an AMD EPYC~7313P (16 cores, 32 threads, 3.0\,GHz) and
125\,GB DDR4 host DRAM.

\textbf{Software.}
Ubuntu~24.04, CUDA~12.9 (Driver~580.105.08), GCC~13.3,
compiled with \texttt{nvcc -O3 -arch=sm\_90}.
The \sysname version under test is main@2026-02;
baseline versions are
WarpCore~v1.3.1, BGHT~v0.1, cuCollections~v0.1.0-ea,
and BP2HT (BGHT main@2026-02).

\looseness=-1
\textbf{Workload parameters.}
Unless stated otherwise, keys are \texttt{uint64\_t}, values are
\texttt{float32}$\times$\texttt{dim}, batch size is 1\,M
key--value pairs per operation, and the eviction strategy is LRU.
We vary embedding dimension
(\texttt{dim}\,$\in$\,$\{$8,\,32,\,64$\}$) and load factor
($\lambda$\,$\in$\,$\{$0.50,\,0.75,\,1.00$\}$).

\textbf{Configurations.}
Table~\ref{tab:configs} summarizes configurations A--C (pure HBM) and D (HBM+HMEM via \texttt{cudaHostAllocMapped}).

\begin{table}[t]
  \caption{Benchmark configurations.  All configurations use a single
  NVIDIA H100 NVL GPU.  Capacity is in millions of key--value pairs (M-KV).}
  \label{tab:configs}
  \centering
  \footnotesize
  \begin{tabular}{clrrrl}
    \toprule
    Config & dim & Capacity & HBM & HMEM & Mode \\
    \midrule
    A & 8  & 128\,M & 4\,GB  & --- & Pure HBM \\
    B & 32 & 128\,M & 16\,GB & --- & Pure HBM \\
    C & 64 & 64\,M  & 16\,GB & --- & Pure HBM \\
    D & 64 & 128\,M & 16\,GB & 16\,GB & HBM+HMEM \\
    \bottomrule
  \end{tabular}
\end{table}

\textbf{Memory overhead.}
Per-entry metadata (key 8\,B, digest 1\,B, score 8\,B) totals 17\,B; values contribute 32--256\,B depending on dimension (6--35\% overhead).

\textbf{Statistical methodology.}
Each value is the median of 5~cold-start runs (coefficient of variation CV${<}3\%$ for reads, ${<}5\%$ for writes); error bars are smaller than plot markers and omitted.

\textbf{Baselines.}
\label{par:baselines}
\looseness=-1
We compare against four dictionary-semantic GPU hash tables compiled with identical CUDA/driver on the same H100 NVL:
WarpCore~\cite{junger2020warpcore}, BGHT~\cite{awad2022better}, cuCollections~\cite{nvidia2020cucollections}, and BP2HT~\cite{awad2022better}.
cuCollections, BGHT, and BP2HT use \emph{(key,\,index) indirection}; all measurements include end-to-end value gather/scatter.\footnote{WarpSpeed~\cite{mccoy2025warpspeed} (scalar-only), FBGEMM TBE~\cite{fbgemm2019} (fused operator), and MemcachedGPU~\cite{hetherington2015memcachedgpu} (network service) are excluded due to incompatible API semantics. Hive~\cite{polak2025hive} (arXiv 2025, dictionary semantics with bounded cuckoo relocation) was not included because its source code was not publicly available at the time of evaluation.}

\subsection{Exp~\#1: Load Factor Analysis}
\label{ssec:exp-lf}

\looseness=-1
Cache semantics deliver stable throughput across the \emph{entire} load factor range,
including $\lambda{=}1.0$---a regime where all dictionary-semantic baselines degrade or fail.

\begin{figure}[t]
  \centering
  \includegraphics[width=\columnwidth]{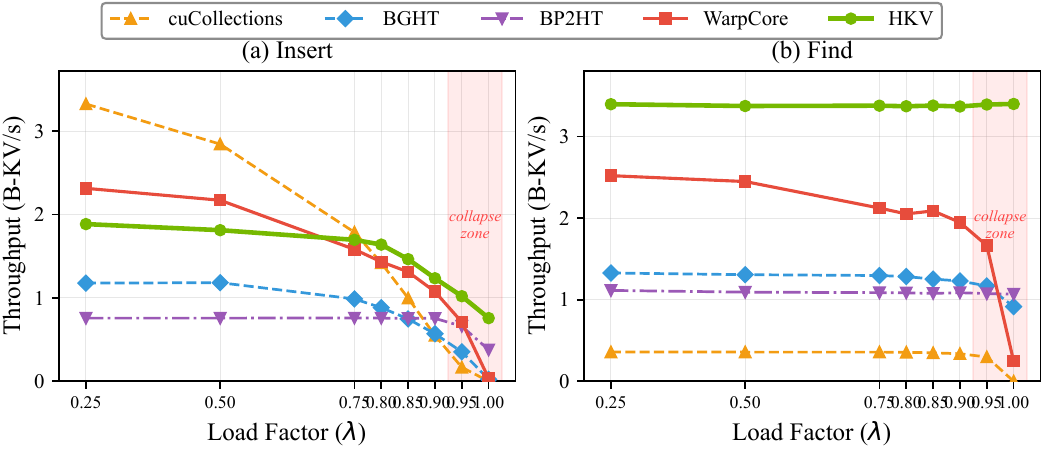}
  \caption{Insert (a) and find (b) throughput vs.\ load factor
    ($\lambda{=}0.25$--$1.00$, dim=32, batch=1\,M).
    \sysname (LRU eviction, cache semantics) maintains near-constant
    \texttt{find} throughput (${\sim}$3.4\,B-KV/s) across the entire
    $\lambda$ range, including $\lambda{=}1.0$, while resolving
    full-bucket inserts in place with no rehash or capacity-induced
    failure.
    Dictionary-semantic baselines degrade monotonically; WarpCore,
    BGHT, and cuCollections collapse beyond $\lambda{\approx}0.95$
    (shaded region).  BP2HT insert throughput is corrected to
    count only successful insertions---at $\lambda{=}1.0$, only
    48\% of inserts succeed.
    Note: cuCollections, BGHT, and BP2HT use (key,\,index)
    indirection as described in \S\ref{ssec:exp-setup}.
    CV${<}$3\% for all data points (5~runs); error bars omitted
    as they are smaller than markers.  (Exp~\#1)}
  \label{fig:lf-degradation}
\end{figure}

\looseness=-1
\textbf{Cross-system comparison.}
Figure~\ref{fig:lf-degradation} plots insert and find throughput
as a function of~$\lambda$ for \sysname and four baselines
(WarpCore, BGHT, cuCollections, BP2HT), all on the same
H100~NVL with dim = 32, batch = 1\,M.

\looseness=-1
\sysname's \texttt{find} throughput varies by less than 1.0\%
across $\lambda{=}0.25$--$1.00$ (3.37--3.40\,B-KV/s), and
every insert is resolved in place with no rehash or failure.
All baselines degrade monotonically:
\texttt{find} drops 90\% (WarpCore), 31\% (BGHT), and 4\% (BP2HT); cuCollections collapses to zero.
Beyond $\lambda{\approx}0.95$ (shaded region), WarpCore,
BGHT, and cuCollections collapse---a
\emph{capability gap}, not merely a performance gap.
BP2HT silently drops insertions: only 48\% succeed at $\lambda{=}1.0$ (effective 0.38\,B-KV/s).
Table~\ref{tab:lf-comparison} quantifies the widening gap.
The load-factor sweep uses uniform-random keys to isolate structural degradation from access-pattern effects; \S\ref{ssec:exp-ablation} (Exp~\#3c) evaluates cache quality under production-like Zipfian skew ($\alpha{=}0.50$--$1.25$).

\begin{table}[t]
  \caption{Find throughput (B-KV/s) at representative load factors.
    H100 NVL, dim=32, batch=1\,M.
    ``${\approx}0$'': near-zero (system stalls).
    (Exp~\#1)}
  \label{tab:lf-comparison}
  \centering\footnotesize
  \begin{tabular}{@{} l ccc cc @{}}
    \toprule
    System & $\lambda{=}0.50$ & $\lambda{=}0.75$ & $\lambda{=}1.00$ & Evict. & Conc. \\
    \midrule
    \textbf{\sysname}  & \textbf{3.37} & \textbf{3.38} & \textbf{3.40} & \textbf{Yes} & \textbf{R/U/I} \\
    WarpCore       & 2.45 & 2.12 & 0.25 & No & LF \\
    BGHT$^\dagger$  & 1.31 & 1.30 & 0.92 & No & LF \\
    cuColl.$^\dagger$ & 0.36 & 0.36 & ${\approx}$0 & No & LF \\
    BP2HT$^\dagger$ & 1.09 & 1.09 & 1.07 & No & LF \\
    \bottomrule
  \end{tabular}

  \smallskip
  {\footnotesize $^\dagger$cuCollections, BGHT, and BP2HT use (key,\,index)
  indirection (\S\ref{ssec:exp-setup}).
  R/U/I = reader/updater/inserter; LF = lock-free CAS.}
\end{table}

At $\lambda{=}0.50$, \sysname \texttt{find} is 1.38--9.43$\times$ faster than all baselines; the gap widens at higher $\lambda$ (Table~\ref{tab:lf-comparison}).

\begin{figure}[t]
  \centering
  \includegraphics[width=\columnwidth]{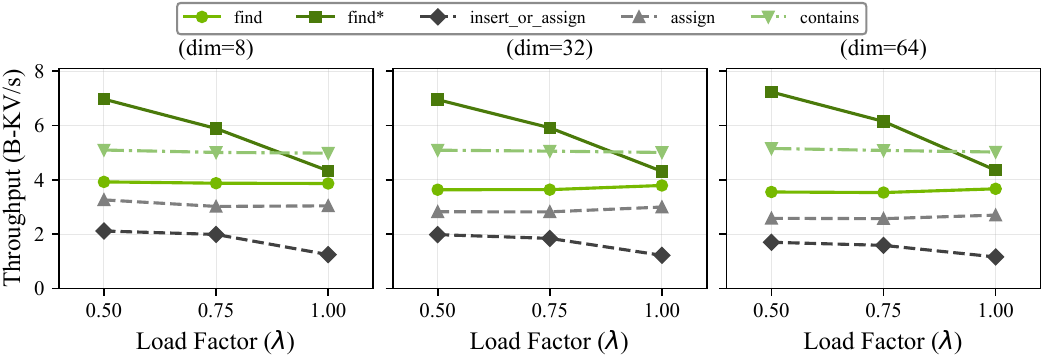}
  \caption{Throughput vs.\ load factor ($\lambda \in \{0.50, 0.75, 1.00\}$)
  in pure HBM mode (Configs~A--C).
  \texttt{find} varies by at most 4.7\% across all configurations (1.3\% at dim\,{=}\,8);
  \texttt{find*} drops ${\sim}$38\% at $\lambda{=}1.00$ due to loss of
  empty-slot early termination;
  \texttt{insert\_or\_assign} shows a bounded 32--41\% decrease at
  $\lambda{=}1.00$ due to eviction scan overhead.  (Exp~\#1)}
  \label{fig:lf-sensitivity}
\end{figure}

\textbf{Per-API sensitivity.}
Figure~\ref{fig:lf-sensitivity} breaks down \sysname's LF sensitivity by API (Configs~A--C).

\looseness=-1
Consistent with the cross-system data, \texttt{find} varies by at most 1.3\% across $\lambda$ (e.g., 3.889--3.941\,B-KV/s at dim=8).  \texttt{find*} degrades by ${\sim}$38\% at $\lambda{=}1.0$ due to loss of empty-slot early termination; \texttt{insert\_or\_assign} decreases 32--41\% from eviction overhead; \texttt{assign} varies by at most 6.7\%, confirming non-structural updates are unaffected by occupancy (at most 6.7\% variation).

\subsection{Exp~\#2: End-to-End Throughput}
\label{ssec:exp-e2e}

\begin{figure}[t]
  \centering
  \includegraphics[width=\columnwidth]{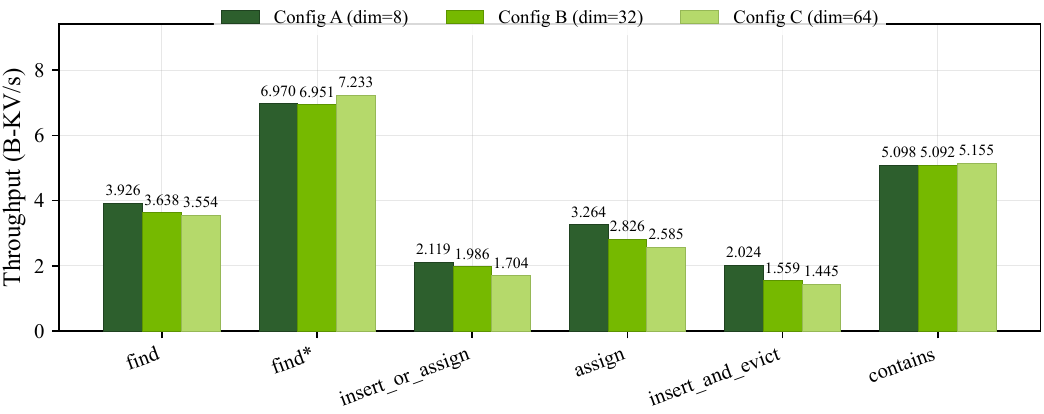}
  \caption{End-to-end throughput of \sysname across all core APIs in
  pure HBM mode (Configurations~A--C, $\lambda{=}0.50$, batch=1\,M~KV/op).
  \texttt{find} ranges from 3.61\,B-KV/s (dim=64) to 3.89\,B-KV/s (dim=8);
  pointer-returning \texttt{find*} is dimension-independent at
  ${\sim}$7.05\,B-KV/s; \texttt{contains} maintains
  ${\sim}$5.10\,B-KV/s across all configurations.  (Exp~\#2)}
  \label{fig:throughput-bar}
\end{figure}

\textbf{Pure HBM throughput.}
\sysname achieves 3.61--3.89\,B-KV/s for \texttt{find} and
${\sim}$7.05\,B-KV/s for the pointer-returning \texttt{find*} across
all embedding dimensions (Figure~\ref{fig:throughput-bar}).

\looseness=-1
\texttt{find} delivers 3.61--3.89\,B-KV/s (7.2\% reduction at dim=64 from value-copy cost); the pointer-returning \texttt{find*} attains ${\sim}$7.05\,B-KV/s independent of dimension, confirming that value copy---not key-side indexing---is the bottleneck.
Write APIs remain in the B-KV/s regime: \texttt{insert\_or\_assign} 1.72--2.13\,B-KV/s, \texttt{insert\_and\_evict} 5--21\% additional overhead, \texttt{assign} 2.59--3.26\,B-KV/s.

\looseness=-1
\textbf{Latency and scaling.}
Per-batch \texttt{find} latency remains stable across load factors (P50\,$\approx$\,0.30\,ms at both $\lambda{=}0.50$ and $1.00$, Config~B, 1{,}000 batches).  Throughput varies ${<}3\%$ across capacities 16\,M--512\,M.

\looseness=-1
\textbf{Hybrid storage impact.}
\label{ssec:exp-hybrid}
Position-based value addressing (\S\ref{ssec:kvsep}) keeps
key-side throughput within 4\% of pure-HBM in hybrid HBM+HMEM mode:
the pointer-returning \texttt{find*} retains 96.0\% throughput
(6.949 vs.\ 7.242\,B-KV/s in pure HBM, Config~D, dim=64, 128\,M
capacity).  Value-copying APIs, by contrast, are bounded by
PCIe: \texttt{find}
drops to 0.172\,B-KV/s ($-$95\%).

\subsection{Exp~\#3: Component Ablation}
\label{ssec:exp-ablation}

We decompose \sysname's throughput to quantify the contribution of
individual design components.

\textbf{Exp~\#3a: Digest contribution.}
We disable digest pre-filtering at compile time and measure \texttt{find} throughput (Table~\ref{tab:digest-ablation}).

\begin{table}[t]
\centering
\footnotesize
\caption{Digest ablation: \texttt{find} throughput (B-KV/s) with and
  without digest pre-filtering.  Speedup = with/without.
  (Exp~\#3a)}
\label{tab:digest-ablation}
\begin{tabular}{clrrr}
\toprule
Config & $\lambda$ & With digest & No digest & Speedup \\
\midrule
A (dim=8)  & 0.50 & 3.376 & 2.052 & 1.65$\times$ \\
B (dim=32) & 0.50 & 3.321 & 1.929 & 1.72$\times$ \\
C (dim=64) & 0.50 & 3.157 & 1.685 & 1.87$\times$ \\
A (dim=8)  & 1.00 & 4.768 & 1.825 & 2.61$\times$ \\
B (dim=32) & 1.00 & 4.769 & 1.828 & 2.61$\times$ \\
C (dim=64) & 1.00 & 4.769 & 1.836 & 2.60$\times$ \\
\bottomrule
\end{tabular}
\end{table}

\noindent At $\lambda{=}0.50$, digest filtering yields 1.65--1.87$\times$ speedup; at $\lambda{=}1.00$ the advantage grows to 2.60--2.61$\times$ because without digest every miss compares all 128 keys (throughput converges to ${\sim}$1.83\,B-KV/s regardless of dimension).

\looseness=-1
\textbf{Exp~\#3b: Eviction overhead.}
Comparing \texttt{insert\_or\_assign} at $\lambda{=}0.50$ (no eviction) vs.\ $\lambda{=}1.00$ (every insert evicts), overhead is 32--41\% (dim=8: 41\%, dim=32: 39\%, dim=64: 32\%)---bounded because the eviction scan always processes exactly one 128-slot bucket.

\looseness=-1
\textbf{Exp~\#3c: Cache quality under production-like access patterns.}
Cache hit rate measures the fraction of lookups that find their target in the table during sustained online ingestion---a runtime metric complementary to top-$N$ retention (\S\ref{ssec:exp-bucket}), which measures the quality of the final table state.
Production recommendation models exhibit power-law embedding access with Zipfian skew $\alpha{\approx}0.99$~\cite{naumov2019dlrm,wang2022hugectr}.
Table~\ref{tab:cache-quality} sweeps $\alpha$ across all five shipped
\texttt{ScoreFunctor} specializations (Config~B, $\lambda{=}1.0$).
At production-typical $\alpha{=}0.99$, LFU achieves 88.3\% (+4.4\,pp over LRU); at $\alpha{\geq}1.25$ all converge to ${\sim}99.4$\% as the hot-set shrinks below capacity.
All policies deliver comparable throughput (2.97--3.04\,B-KV/s), confirming that the contribution is the shared in-line upsert mechanism rather than policy-count breadth.

\begin{table}[t]
\centering
\footnotesize
\caption{Cache hit rate (\%) by scoring policy and Zipfian $\alpha$
  (Config~B, $\lambda{=}1.0$, single-bucket, capacity\,${=}$\,128\,M).
  The $\alpha$ sweep is used as a sensitivity study around
  production-like power-law skew; see text.
  (Exp~\#3c)}
\label{tab:cache-quality}
\begin{tabular}{l rrrr}
\toprule
Policy & $\alpha{=}0.50$ & $\alpha{=}0.75$ & $\alpha{=}0.99$ & $\alpha{=}1.25$ \\
\midrule
kLru        & 18.6 & 43.0 & 83.9 & 99.4 \\
kLfu        & \textbf{31.4} & \textbf{55.7} & \textbf{88.3} & \textbf{99.7} \\
kEpochLru   & 18.6 & 43.0 & 83.9 & 99.4 \\
kEpochLfu   & 18.6 & 43.0 & 83.9 & 99.4 \\
kCustomized & 18.6 & 43.0 & 83.9 & 99.4 \\
\bottomrule
\end{tabular}
\end{table}

\looseness=-1
\textbf{Exp~\#3d: Admission control.}
The score-based admission gate (Algorithm~\ref{alg:upsert}, line~10) rejects keys scoring below the bucket minimum.
A burst experiment (16\,M-slot table at $\lambda{=}0.96$, 4\,M injected keys) isolates the gate's contribution (Table~\ref{tab:admission}).

\begin{table}[t]
\centering
\footnotesize
\caption{Admission control ablation: hit-rate change after a 4\,M-key
  burst injection into a 16\,M-slot table at $\lambda{=}0.96$.
  (Exp~\#3d)}
\label{tab:admission}
\begin{tabular}{lcc}
\toprule
Burst score & Admitted & $\Delta$ hit rate \\
\midrule
Low ($s{=}1$) & 0\% (fully rejected) & $+$0.00\,pp \\
High ($s{=}10^9$) & 100\% (displaces originals) & $-$21.48\,pp \\
\bottomrule
\end{tabular}
\end{table}


\looseness=-1
\textbf{Exp~\#3e: Concurrency ablation.}
We replace triple-group (\S\ref{ssec:triplegroup}) with a conventional R/W lock and measure seven workload mixes ($\lambda{=}0.75$, dim\,{=}\,16, 64\,K keys/batch, 200 batches/thread).
As updaters scale from 1 to 10, triple-group reaches 2.569\,B-KV/s while R/W lock anti-scales to 0.535\,B-KV/s ($4.80\times$).
Under mixed workloads the gap narrows but remains significant: update-heavy (4F/5U/1I) $3.21\times$, insert-heavy (4F/2U/4I) $1.20\times$; read-heavy (8F/1U/1I) shows near-parity ($1.03\times$) because a single updater rarely contends.

\looseness=-1
Table~\ref{tab:codesign-ablation} summarizes the system-level impact of
removing each component, demonstrating that the four innovations form a
non-decomposable design: each removal breaks a downstream emergent property.

\begin{table}[t]
\caption{Co-design ablation: removing any component breaks a system-level property.  H100~NVL (\S\ref{ssec:exp-e2e}--\S\ref{ssec:exp-bucket}).}
\label{tab:codesign-ablation}
\centering\footnotesize
\begin{tabular}{@{}lll@{}}
\toprule
Removed & Broken Property & Measured Impact \\
\midrule
Eviction & Insert fails at $\lambda{>}0.66$ & Cannot sustain $\lambda{=}1.0$ \\
Dual-bucket & Early eviction, lower retention & $\lambda_{\text{1st}}$: .98$\to$.63; top-$N$: 99.4$\to$95.4\% \\
Triple-group & Updates serialize & $4.80\times$ slower \\
Single-bucket & Miss needs ${\geq}$2 loads & 1-txn definitive miss lost \\
\bottomrule
\end{tabular}
\end{table}

\subsection{Exp~\#4: Single-Bucket vs.\ Dual-Bucket}
\label{ssec:exp-bucket}


\begin{table}[t]
  \caption{Single-bucket vs.\ dual-bucket mode under sustained
  Zipfian ingestion ($\alpha{=}0.99$, capacity${=}128$\,M,
  $b{=}128$).
  Top-$N$ retention measures the fraction of the ideal
  highest-scored $N$ keys ($N{=}$\,capacity) present in the
  table after $5{\times}$\,capacity steady-state inserts at
  $\lambda{=}1.0$.
  (Exp~\#4)}
  \label{tab:single-vs-dual}
  \centering
  \footnotesize
  \begin{tabular}{lccc}
    \toprule
    \textbf{Metric} & \textbf{Single} & \textbf{Dual} & \textbf{$\Delta$} \\
    \midrule
    First-eviction $\lambda$      & 0.633  & 0.977  & +54.3\% \\
    Top-$N$ score retention       & 95.39\% & 99.44\% & +4.05\,pp \\
    Cache hit ratio               & 83.88\% & 84.02\% & +0.14\,pp \\
    \bottomrule
  \end{tabular}
\end{table}

\looseness=-1
Dual-bucket selection delays first eviction from $\lambda{=}0.633$ to
$\lambda{=}0.977$ (+54.3\%) and raises top-$N$ score retention to
99.44\% (Table~\ref{tab:single-vs-dual}).
Results compare the dual-bucket GPU kernel (\S\ref{ssec:dualbucket}) against the default single-bucket mode.

\looseness=-1
\textbf{Throughput.}
Dual-bucket mode delivers comparable or higher throughput than single-bucket across all load factors.
At $\lambda{=}0.50$, dual-bucket \texttt{find} throughput is ${\sim}5\%$ higher than single-bucket---two-choice placement distributes keys more evenly across buckets, reducing per-bucket occupancy and shortening the average digest-scan path---and \texttt{insert\_or\_assign} throughput is within $1\%$.
At $\lambda{=}1.0$, the advantage widens: dual-bucket \texttt{insert\_or\_assign} attains $1.64{\times}$ higher throughput than single-bucket because single-bucket suffers from premature eviction overhead starting at $\lambda{\approx}0.66$, while dual-bucket delays eviction onset to $\lambda{\approx}0.98$.
Combined with the $54.3\%$ improvement in first-eviction load factor and $4.05$\,pp gain in top-$N$ score retention (Table~\ref{tab:single-vs-dual}), dual-bucket offers strictly superior behavior for memory-constrained deployments with no throughput penalty.

%% file: sections/s6-related.tex

\section{Related Work}
\label{sec:related}

\looseness=-1
\paragraph{GPU Hash Tables under Dictionary Semantics.}
GPU hash tables have evolved from cuckoo-based designs~\cite{alcantara2012building} through Stadium Hashing~\cite{khorasani2015stadium}, warp-cooperative slabs~\cite{ashkiani2018dynamic}, WarpCore~\cite{junger2020warpcore}, cuCollections~\cite{nvidia2020cucollections}, BGHT~\cite{awad2022better}, and WarpSpeed~\cite{mccoy2025warpspeed}.  DyCuckoo~\cite{li2021dycuckoo} addresses dynamic resizing; DACHash~\cite{zhou2021dachash} optimizes GPU hardware cache efficiency; GPHash~\cite{chen2025gphash} targets byte-granularity persistent memory; Hegeman et al.~\cite{hegeman2024compact} adapt compact hashing to GPU buckets but still require multi-bucket probing.  For a comprehensive survey of GPU hashing techniques, see Lessley and Childs~\cite{lessley2020gpu}.
These systems differ in collision resolution and hardware optimization, but they retain dictionary semantics: every inserted key must be preserved, so a full table triggers rehashing or insertion failure.

\looseness=-1
\paragraph{Bounded-Probe, Two-Choice, and Fingerprinted Miss Filtering.}
The power-of-two-choices paradigm~\cite{azar1994balanced,mitzenmacher2001power,seznec1993skewed} proves that choosing the less loaded of two bins exponentially reduces maximum load; Panigrahy~\cite{panigrahy2005efficient}, Walzer et al.~\cite{walzer2023load}, and BP2HT~\cite{awad2022better} connect this lineage to bucketized and GPU hash tables (BP2HT uses static load-based selection and fails when both buckets are full), while bounded-probe designs---Hopscotch~\cite{herlihy2008hopscotch}, bucketized cuckoo~\cite{pagh2004cuckoo,dietzfelbinger2007balanced}, Horton Tables~\cite{breslow2016horton}---still rely on overflow mechanisms under dictionary semantics.  CPU hash tables---SwissTable~\cite{swisstable2017}, F14~\cite{f14folly2019}, BBC~\cite{bother2023bbc}---use inline fingerprints but require multi-group traversal for misses.  On the miss path specifically, MemC3~\cite{fan2013memc3} probes two buckets of 4~slots with no GPU support; MemcachedGPU~\cite{hetherington2015memcachedgpu} embeds 8-bit hashes within ${\ge}$20\,B per-entry headers, requiring up to 16~candidate slots across multiple cache-line loads per miss; BP2HT~\cite{awad2022better} and Hive~\cite{polak2025hive} (arXiv, 2025) use multi-bucket designs without digest arrays, incurring ${\ge}$2 bucket reads per miss.  In contrast, \sysname combines single-bucket confinement with a contiguous digest array, so each bucket check is definitive and costs one fixed cache-line load.

\looseness=-1
\paragraph{Cache Replacement and Eviction-Aware Data Structures.}
Belady's optimal algorithm~\cite{belady1966study} provides the offline lower bound; practical online policies trade optimality for $O(1)$ overhead.
RRIP~\cite{jaleel2010rrip} demonstrates that prediction-based replacement outperforms recency-only policies in hardware caches---a principle \sysname applies at the hash-table library level through a shared score-computation interface.
CPHash~\cite{metreveli2011cphash} introduces LRU-aware hash tables; Friedman and Mozes~\cite{friedman2021limited} show that limited associativity improves throughput with marginal hit-ratio impact; Cavast~\cite{wu2020cavast} optimizes data layout at the OS page level for CPU LLC efficiency; \sysname operates at the hash table bucket level for GPU L1 cache-line alignment.
MemC3~\cite{fan2013memc3} co-designs cuckoo hashing with tags and CLOCK eviction; CacheLib~\cite{berg2020cachelib} provides production-grade pluggable eviction via linked-list-based MMContainers decoupled from the hash index.
TinyLFU~\cite{einziger2017tinylfu} proposes a lightweight admission gate that \sysname adapts to per-bucket score-based admission (\S\ref{ssec:eviction}).
HashPipe~\cite{sivaraman2017hashpipe} embeds count-based eviction into a data-plane pipeline of single-entry hash stages; \sysname shares the in-path eviction principle but operates on 128-slot GPU buckets with pluggable scoring and general key--value semantics.
Unlike all of these, \sysname fuses eviction scoring into the insert CAS path with no separate eviction queue, eviction data structure, or admission filter---the hash table \emph{is} the cache.

\looseness=-1
\paragraph{Concurrent Hash Tables.}
Li et al.~\cite{li2014concurrent} achieve fine-grained concurrent cuckoo hashing with optimistic locking; RCU hash tables~\cite{triplett2011rcu} define reader/writer/resizer roles; Fatourou et al.~\cite{fatourou2018waitfree} separate per-bucket mutation from directory resizing---Fatourou's update role bundles \emph{all} mutations (insert, delete, value change) into one role via per-bucket PSim instances, whereas \sysname further decomposes mutations: non-structural value/score updates run concurrently with lookups, and only structural inserts (which may trigger eviction) require exclusive access, constituting ${\sim}$10\% of mixed-workload operations.
Bw-tree~\cite{levandoski2013bwtree} formalizes structure modification operations; oneTBB's \texttt{concurrent\_hash\_map}~\cite{malakhov2015onetbb} provides per-bucket reader/writer accessors.
Liu et al.~\cite{liuzspear2014dynamic} present a three-phase freeze-migrate pattern for nonblocking hash table resizing; no prior GPU hash table separates value/score updates from insert/evict structure changes.
\sysname adapts the structural-vs.-non-structural split to GPU hash tables, separating value/score updates (parallelizable) from inserts/evictions (exclusive), with a CPU--GPU dual-layer lock and role-isolated kernels absent from prior CPU-only designs.

\looseness=-1
\paragraph{Key-Value Separation and Tiered Storage.}
Key-value separation was pioneered by WiscKey~\cite{lu2016wisckey} and HashKV~\cite{chan2018hashkv} in the SSD/LSM-tree context; FAWN-DS~\cite{andersen2009fawnds}, SILT~\cite{lim2011silt}, and FASTER~\cite{chandramouli2018faster} employ compact in-memory hash indices pointing to storage-resident data.
MICA~\cite{lim2014mica} pioneered holistic KV store design with circular-log value storage and parallel hash-index partitioning;
MemcachedGPU~\cite{hetherington2015memcachedgpu} and Mega-KV~\cite{zhang2015megakv} represent early GPU key-value separation but store per-entry value pointers and require CPU threads on the critical path (MemcachedGPU for SET processing from CPU-resident slabs, Mega-KV for allocation and eviction scheduling).
\sysname introduces \emph{position-based value addressing}: keys, digests, and scores reside in HBM while values overflow to HMEM via \texttt{cudaHostAllocMapped}---each slot's key and value share the same array index, eliminating per-entry value pointers and keeping the CPU off the critical path via zero-copy mapped pointers.

\looseness=-1
\paragraph{Embedding Storage for Recommendations.}
Deep learning recommendation models~\cite{naumov2019dlrm} have driven specialized embedding storage.
Meta's FBGEMM~\cite{fbgemm2019} is \sysname's closest industrial analogue, providing GPU set-associative LRU/LFU cache with 8--32 ways (version-dependent).
The key architectural difference is \emph{where eviction decisions are made}: FBGEMM decouples eviction into a multi-kernel pipeline with separate data structures, whereas \sysname embeds score comparison, admission control, and victim replacement directly into the bucket-local insert path.
NVIDIA's cuEmbed~\cite{anderson2025cuembed} and nv-embedding-cache~\cite{nvidia2026nvembcache} independently address the same problem with an 8-way set-associative design, further validating the importance of GPU embedding caching.
Fleche~\cite{xie2022fleche} and UGACHE~\cite{song2026ugache} provide GPU embedding caching with approximate LRU and static placement respectively; Catalyst~\cite{wang2023catalyst} automates caching decisions via learned cost models; LCR~\cite{chen2025lcr} applies learned replacement at the framework level.
\sysname serves as a \emph{low-level GPU hash table primitive} with in-kernel eviction; higher-level systems can layer framework-specific policy on top.

%% file: sections/s7-conclusion.tex

\section{Conclusion}
\label{sec:conclusion}

\looseness=-2
We have presented \sysname, the first GPU hash table library whose normal full-capacity operating contract is cache-semantic: eviction is a first-class, policy-driven operation, not an error or maintenance event.
Four co-designed mechanisms deliver three core system-level properties:
cache-line-aligned 128-slot buckets with contiguous digest arrays achieve \emph{definitive per-bucket miss in one memory transaction} (\S\ref{ssec:bucket});
in-line score-driven upsert combined with score-based dual-bucket selection resolves full-bucket inserts in place via eviction or admission rejection while improving retention at $\lambda{=}1.0$ (\S\ref{ssec:eviction}, \S\ref{ssec:dualbucket});
and the triple-group concurrency protocol preserves mixed-workload throughput by separating structural from non-structural GPU writes (\S\ref{ssec:triplegroup}).
As a scaling enabler, tiered key-value separation extends the same design beyond HBM while keeping key-side processing on GPU (\S\ref{ssec:kvsep}).
Together, these properties enable \sysname to sustain stable billion-scale throughput at load factor~1.0 on an NVIDIA H100 NVL GPU---a combination unattainable by any single prior component.
Open-source integrations in NVIDIA Merlin HugeCTR~\cite{wang2022hugectr}, TFRA~\cite{tfra2020}, and NVIDIA RecSys Examples~\cite{nvidia2024recsys} provide a practical deployment path for GPU-accelerated embedding storage.
\textbf{Limitations and future work.}
Reader--updater mutual exclusion trades atomicity for simplicity (relaxable with single-word-atomic values); dual-bucket mode covers \texttt{insert\_or\_assign} and \texttt{find} while the single-bucket path supports the full API; multi-GPU sharding is delegated to application code by design.
Looking ahead, wider GPU cache lines would enable larger buckets; SSD/GDS tiering and dynamic rehash remain open extensions.
More broadly, \sysname establishes cache semantics as a practical design paradigm for GPU-resident data structures---one that trades the dictionary invariant for sustained full-capacity operation under continuous ingestion.